%% file: main.tex
\documentclass[twocolumn]{autart}    % Enable this line and disable the 
\input{tikz_packages}

\input{colors}

\usepackage{graphicx}          % Include this line if your 
\usepackage{booktabs}

\usepackage{mdframed}

\usepackage{cite}

\usepackage{subcaption}

\usepackage{amsmath}
\usepackage{amssymb}
\usepackage{mathtools}
\usepackage{thmtools}

\theoremstyle{plain}
\newtheorem{theorem}{Theorem}

\newtheorem{corollary}{Corollary}

\theoremstyle{definition}
\newtheorem{definition}{Definition}
\newtheorem{assumption}{Assumption}
\newtheorem{problem}{Problem}

\theoremstyle{remark}
\newtheorem{remark}{Remark}
\newtheorem{example}{Example}

\renewcommand{\thmcontinues}[1]{Continued}

\DeclareMathAlphabet{\mathbbold}{U}{bbold}{m}{n}

\usepackage{xcolor}

\usepackage{float}

\usepackage{algpseudocode,algorithm,algorithmicx}

\DeclareMathOperator*{\argmax}{argmax}

\begin{document}

\begin{frontmatter}

\title{Temporal Logic Control of Nonlinear Stochastic Systems \\ with Online Performance Optimization
% \thanksref{footnoteinfo}
} % Title, preferably not more 

% \thanks[footnoteinfo]{This paper was not presented at any IFAC 
% meeting. Corresponding author M.~T.~Cicero. Tel. +XXXIX-VI-mmmxxi. 
% Fax +XXXIX-VI-mmmxxv.}

\author[a]{Alessandro Riccardi}\ead{a.riccardi@tudelft.nl},   
\author[b,c]{Thom Badings}\ead{thom.badings@cs.rwth-aachen.de},            
\author[a,d]{Luca Laurenti}\ead{l.laurenti@tudelft.nl},
\author[b]{Alessandro Abate}\ead{alessandro.abate@cs.ox.ac.uk},~and~\author[a]{Bart De Schutter}\ead{b.deschutter@tudelft.nl}.

\address[a]{Delft University of Technology, The Netherlands}  
\address[b]{University of Oxford, United Kingdom}
\address[c]{RWTH Aachen University, Germany}  
\address[d]{AI4I, Italy}

                             % keyword list or with the 
                                          % help of the Automatica 
                                          % keyword wizard

\begin{abstract}                          % Abstract of not more than 200 words.
The deployment of autonomous systems in safety-critical environments requires control policies that guarantee satisfaction of complex control specifications. These systems are commonly modeled as nonlinear discrete-time stochastic systems. A~popular approach to computing a policy that provably satisfies a complex control specification is to construct a finite-state abstraction, often represented as a Markov decision process (MDP) with intervals of transition probabilities, i.e., an interval MDP (IMDP). However, existing abstraction techniques compute a \emph{single policy}, thus leaving no room for online cost or performance optimization, e.g., of energy consumption. To overcome this limitation, we propose a novel IMDP abstraction technique that yields a \emph{set of policies}, each of which satisfies the control specification with a certain minimum probability. We can thus use any online control algorithm to search through this set of verified policies while retaining the guaranteed satisfaction probability of the entire policy set. In particular, we employ model predictive control (MPC) to minimize a desired cost function that is independent of the control specification considered in the abstraction. Our experiments demonstrate that our approach yields better control performance than state-of-the-art single-policy abstraction techniques, with a small degradation of the guarantees.
\end{abstract}

\end{frontmatter}

\section{Introduction}

Modern autonomous systems, such as unmanned drones and robotic systems, can be generally described by \emph{nonlinear stochastic dynamical systems}~\cite{kumar2015stochastic,mesbah2016stochastic}.
These systems must meet stringent requirements across multiple facets of their behavior.
First, systems must perform complex tasks expressed as \emph{logical specifications}~\cite{BaierKatoen08,belta2017formal}, e.g., a robot moving an object from location A to B while avoiding obstacles. 
Second, systems must optimize for other key performance indicators modeled as a \emph{cost} (or \emph{reward}) \emph{function}~\cite{aastrom2012introduction,fleming2012deterministic}, such as minimizing the energy consumption or time to complete a task. 
To deploy autonomous systems in real-world environments, automated techniques for synthesizing control policies that provably meet the first type of requirement, whilst optimizing the latter, have now become imperative.

Existing policy synthesis techniques, however, largely focus on \emph{either} satisfying logical specifications or optimizing a cost function. 
For example, \emph{abstraction-based approaches}~\cite{LSAZ21,DBLP:journals/tac/LahijanianAB15,DBLP:journals/jair/BadingsRAPPSJ23,DBLP:journals/tac/HaesaertS21} arguably are the state of the art for computing policies that provably satisfy complex logical specifications---e.g., in \emph{linear temporal logic} (LTL)~\cite{DBLP:conf/focs/Pnueli77}---under stochastic and nonlinear dynamics. 
However, the resulting policy is computed offline and generally cannot be modified without losing its correctness guarantees. 
On the other hand, common control methodologies such as \emph{model predictive control} (MPC)~\cite{mesbah2016stochastic,rawlings_ModelPredictiveControl_2017} excel at online cost minimization but generally cannot provide guarantees about the probability of satisfying complex specifications (such as those expressed as logical specifications). 
As a result, existing methodologies are unable to provide guarantees on the satisfaction of logical specifications, while also optimizing a desired cost function.

In this paper, we address this gap by proposing a novel integration of abstraction-based policy synthesis techniques with online MPC-based optimization. 
Focusing on stochastic, nonlinear systems evolving in discrete time, we consider the following problem:
\vspace{0.5em}
\begin{mdframed}[
nobreak=true, innerleftmargin=8pt, innerrightmargin=8pt]
\textbf{Problem.} Given a discrete-time stochastic dynamical system, a logical specification to be satisfied with a desired probability threshold $\lambda$, and a cost function $J$, compute a policy such that its probability of satisfying the specification is at least $\lambda$, and such that it minimizes the cost function~$J$.
\end{mdframed}
\vspace{-0.5em}

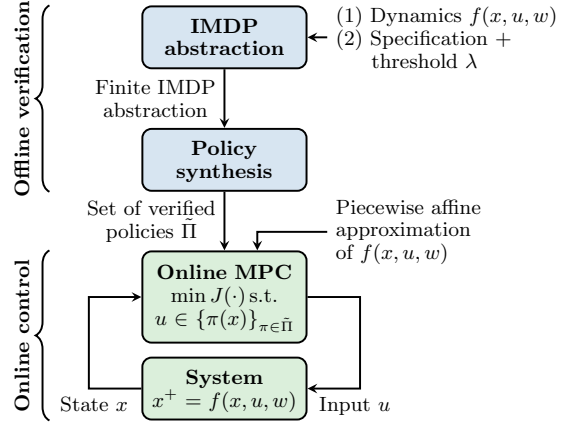
\begin{figure}[t!]%
    \centering
    \scalebox{0.96}{
    \input{overview}%
    }%
    \caption{Our framework leverages formal abstractions to compute a set of certified policies $\tilde\Pi$, together with online MPC to optimize a cost function within this certified set. 
    }%
    \label{fig:overview}%
\end{figure}

We split this problem into two parts, as shown in Fig.~\ref{fig:overview}.
Offline, we construct a finite-state abstraction of the dynamical system, represented as a \emph{Markov decision process} (MDP) with intervals of transition probabilities---called an \emph{interval MDP} (IMDP)~\cite{DBLP:journals/ior/NilimG05,DBLP:journals/mor/Iyengar05}.
As a novel feature, our abstraction procedure allows us to compute a \emph{set of policies} $\tilde{\Pi}$ for the original system, each of which satisfies the logical specification within the probability threshold $\lambda$. 
(By contrast, existing abstraction procedures yield only a \emph{single} policy.)
Intuitively, this set of policies allows a subset of control inputs in every state, also known as a \emph{permissive policy}.
Online, we implement an MPC algorithm that minimizes the cost function $J$ over this set of policies $\tilde{\Pi}$. 
As a result, the closed-loop dynamical system is guaranteed to satisfy the logical specification with probability at least $\lambda$, while also minimizing the cost function $J$. 
We now discuss the key features of our abstraction and MPC~algorithm, respectively.

\textbf{Offline abstraction\,\,} 
Existing approaches construct IMDPs in which each abstract action corresponds with a \emph{single} control input for the dynamical system.
As a result, an abstract policy determines \emph{all} control choices for the dynamical system, rendering these techniques incompatible with online optimization. 
As a novel feature, we associate each abstract action with a \emph{set of control inputs} for the dynamical system.
Thus, every abstract policy corresponds with a \emph{set of policies} for the dynamical system, each of which satisfies the specification with probability at least $\lambda$.
We prove the correctness of this novel abstraction approach by showing that our abstraction induces a variant of a \emph{probabilistic alternating simulation relation} (PASR)~\cite{DBLP:journals/njc/SegalaL95,Badings2025CDC}.

\textbf{Online control\,\,} 
The MPC controller minimizes the cost function $J$ while constraining control inputs to the subset allowed by the set of policies obtained from the abstraction.
Due to the state-space discretization, these policies are piecewise constant functions of the abstract state, which, together with the nonlinear dynamics, render the MPC problem nonconvex.
Thus, we leverage reformulations from MPC for hybrid systems to obtain a \emph{piecewise affine} approximation of the dynamics \cite{sontag_NonlinearRegulationPiecewise_1981}.
We formulate the resulting MPC problem as a mixed integer quadratic program (MIQP)~\cite{bemporad_ControlSystemsIntegrating_1999}.
Crucially, despite these approximations, the online MPC solution preserves the certified lower bound $\lambda$ on the satisfaction probability obtained via the abstraction.
Indeed, even if the MIQP is infeasible (which may occasionally occur due to the model approximation used by the MPC controller), choosing \emph{any} input consistent with the set of policies obtained from the abstraction ensures that the satisfaction probability threshold of $\lambda$ is preserved.

\textbf{Contributions\,\,}
Our main contributions are as follows:%
\begin{itemize}
    \item \textbf{Theoretic:} We extend existing notions of simulation relations for IMDP abstractions to associate each abstract action with a set of control inputs for the original dynamical system: this newly empowers abstractions to be compatible with online control.
    \item \textbf{Algorithmic:} We develop a tailored MPC scheme that optimizes a given cost function while preserving the lower bound on the satisfaction probability obtained from the abstraction.
    \item \textbf{Empirical:} We evaluate our framework on existing benchmarks~\cite{towers_GymnasiumStandardInterface_2025,moore1990efficient,Badings2025CDC} and compare against a vanilla IMDP abstraction technique with no cost optimization, as in, e.g.,~\cite{DBLP:journals/tac/LahijanianAB15}.
    The experiments confirm that our framework effectively improves the cost function $J$ at a marginal reduction of the lower bound $\lambda$ on the probability of satisfying the logical specification.
\end{itemize}

\textbf{Structure of the paper\,\,}
After the related work in Sec.~\ref{sec:related}, we formalize the problem in Sec.~\ref{sec:Problem} and introduce IMDPs in Sec.~\ref{sec:IMDPs}.
We then present our novel behavioral relation based on sets of control inputs in Sec.~\ref{sec:Interface}, which we embed in an IMDP abstraction algorithm in Sec.~\ref{sec:IMDP_abstraction}.
Thereafter, Sec.~\ref{sec:MPC} integrates this IMDP abstraction into an MPC scheme for online cost optimization.
Finally, we empirically evaluate our abstraction framework in Sec.~\ref{sec:Experiments}.

\section{Related work} \label{sec:related}
Policy synthesis for stochastic dynamical systems has largely been addressed via two approaches~\cite{LSAZ21}.
The first (which we adapt in this work) is to construct a---typically finite---abstraction and to synthesize a policy on this abstract model~\cite{Abate2008probabilisticSystems,DBLP:journals/tac/LahijanianAB15}.
Under an appropriate relation (e.g., a simulation or bisimulation relation), bounds on the probability of satisfying a control specification carry over from the abstraction to the original system.
Abstractions are commonly represented as MDPs~\cite{DBLP:journals/tac/HaesaertS21,DBLP:conf/hybrid/HuijgevoortSSH23,DBLP:conf/hybrid/Lavaei24}, IMDPs~\cite{DBLP:journals/tac/LahijanianAB15,DBLP:journals/jair/BadingsRAPPSJ23,DBLP:conf/hybrid/CauchiLLAKC19}, or variants with more expressive forms of uncertainty~\cite{DBLP:conf/hybrid/MathiesenHL25,DBLP:journals/corr/abs-2507-02213}.
Recently, abstractions constructed from data have also gained significant interest~\cite{DBLP:conf/l4dc/GraciaLMAL25,DBLP:conf/hybrid/JacksonLFL21,DBLP:conf/l4dc/NazeriBSA25,DBLP:journals/csysl/LavaeiSFZ23}.

The second approach to policy synthesis for stochastic systems is to find a \emph{certificate function} that implies the satisfaction of a specification.
Certificates have been developed for stability~\cite{DBLP:journals/tac/QinCA20}, safety~\cite{DBLP:journals/tac/PrajnaJP07,DBLP:journals/automatica/Clark21,DBLP:journals/tac/JagtapSZ21}, reach-avoidance~\cite{DBLP:conf/cav/BadingsKJJ25,DBLP:conf/aaai/ZikelicLHC23}, and recently for general (\emph{$\omega$-regular}) specifications~\cite{DBLP:conf/cav/AbateGR25,DBLP:conf/cav/HenzingerMSZ25}.
Yet, finding certificates is challenging, and developing general methods for computing them remains an ongoing research effort~\cite{dawson_safe_2022}. 

Abstraction and certificate-based approaches are primarily designed to certify or maximize the probability of satisfying a given logical specification, without accounting for a user-defined cost function such as energy consumption or control effort. 
Abstraction techniques use dynamic programming to optimize the probability of satisfying the temporal specification~\cite{LSAZ21}. 
Similarly, certificate-based approaches compute a function (e.g., a control barrier function) that proves the existence of a policy that satisfies a specification with a desired probability~\cite{DBLP:journals/tac/JagtapSZ21}. 
A fundamental limitation of both approaches is that the synthesized policy is determined entirely offline, as part of constructing the abstraction or certificate.
Hence, the policy cannot be modified in response to newly observed system behavior or changing operating conditions, leaving no room for online adaptation or performance improvement without losing the guarantees. 

MPC is among the most popular modern control techniques \cite{rawlings_ModelPredictiveControl_2017,mesbah2016stochastic}.
Given a current state, MPC solves an optimization problem that minimizes a cost function over a finite horizon while accounting for the dynamics and constraints on, e.g., inputs and states.
Current research on MPC focuses on integration with data-based and machine-learning techniques~\cite{morato_DataScienceModel_2024, reiter_SynthesisModelPredictive_2025a}, and on large-scale and distributed applications \cite{scattolini_ArchitecturesDistributedHierarchical_2009,  riccardi_PartitioningTechniquesNoncentralized_2025a}.
In this work, we use MPC for hybrid dynamical systems~\cite{tabuada_VerificationControlHybrid_2009,heemels_EquivalenceHybridDynamical_2001} to connect the abstraction-based policy with online optimization. Specifically, we use logical constraints to select optimization regions, based on the framework for mixed logical dynamical systems~\cite{bemporad_ControlSystemsIntegrating_1999}. Despite the maturity and versatility of MPC, no technique has been developed so far that guarantees the satisfaction of complex specifications, such as temporal logic specifications \cite{LSAZ21}.
For example, for reach-avoid specifications, MPC can be augmented with an external  trajectory planner \cite{siciliano_RoboticsModellingPlanning_2009} that provides a reference for the state. 
However, MPC with trajectory planning in general cannot provide formal guarantees about the satisfaction of temporal logic specifications under stochastic, nonlinear dynamics.
Our approach instead does provide such guarantees, and allows the MPC to choose from a set of control inputs prescribed by the abstraction policy.    

Finally, our setting is conceptually related to \emph{safety filters} or \emph{shields} that monitor a policy at runtime and intervene, when necessary, by modifying its intended action~\cite{DBLP:journals/arcras/HsuHF24,DBLP:journals/arcras/HewingWMZ20,DBLP:journals/automatica/WabersichZ21}.
Safety filters have been synthesized, e.g., as control barrier functions~\cite{DBLP:journals/tac/AmesXGT17,DBLP:journals/tac/Krstic24} and as shields using model checking~\cite{DBLP:conf/aaai/AlshiekhBEKNT18,DBLP:conf/aaai/Carr0JT23}.
Nevertheless, as their name suggests, safety filters only guarantee the satisfaction of control constraints or forward invariance, but not the satisfaction of the richer temporal logic specifications we consider in this paper.
Our method also differs from the stochastic optimal control methodology for constrained MDPs developed in \cite{ni_LearningbasedApproachStochastic_2025}, which uses a learning-based policy gradient approach to ensure safe exploration.

\section{Problem Setup}
\label{sec:Problem}

\textbf{Notation\,\,}
The power set over $X$ is written $2^X$.
For the natural numbers, we write ${\mathbb{N}}_0 \coloneqq \{0\} \cup {\mathbb{N}}$.
A probability space $(\Omega, \mathcal{F}, {\mathbb{P}})$ consists of a sample space $\Omega$, a $\sigma$-algebra $\mathcal{F}$, and a probability measure ${\mathbb{P}} \colon \mathcal{F} \to [0,1]$.
The Borel $\sigma$-algebra over a set $X$ is $\mathcal{B}(X)$.
The set of all distributions over a set $X$ with a $\sigma$-algebra $\mathcal{F}$ is written as $\Delta(X,\mathcal{F})$.
For a finite set $X$, we omit the $\sigma$-algebra and write $\Delta(X)$.

\subsection{Discrete-time stochastic systems}
Consider a discrete-time nonlinear stochastic system, denoted by ${\mathbf{S}}$, described by the difference equation
\begin{align} \label{eq:dynamics}
    {\mathbf{S}} \colon 
    \begin{cases}
        x_{k+1} = f(x_k,u_k,w_k), \enskip \textrm{for} \,\, k = 0,1,\ldots, \\
        \quad x_{0} \sim {\mu_{x_0}},
    \end{cases}
\end{align}
where $x_k\in{\mathcal{X}}\subseteq\mathbb{R}^{n_x}$, $u_k\in{\mathcal{U}}\subseteq\mathbb{R}^{n_u}$, are, respectively, the states and control input at time step $k$ (with state space ${\mathcal{X}}$ and input space ${\mathcal{U}}$), $w_k \in {\mathcal{W}}$ is a stochastic disturbance, ${\mu_{x_0}} \in \Delta({\mathcal{X}}, \mathcal{B}(X))$ is the initial state distribution, and $f\colon{\mathcal{X}}\times{\mathcal{U}}\times{\mathcal{W}} \to {\mathcal{X}}$ is the transition function. 

\begin{assumption}
    \label{assumption:disturbance}
    The disturbance $\{w_{k}\}_{k \in {\mathbb{N}}_0}$ is a stationary process, where each $w_k \sim {\mathbb{P}}$ is an i.i.d. random variable in the probability space $({\mathcal{W}},\mathcal{B}({\mathcal{W}}),{\mathbb{P}})$, where ${\mathbb{P}}$ is absolutely continuous with respect to the Lebesgue measure.
\end{assumption}

The probability measure ${\mathbb{P}}$ being absolutely continuous implies that the probability for $x_{k+1}$ to lie in a zero-volume set is zero, which will be a desirable property when generating abstractions (see Footnote~\ref{fn:partition} in Sec.~\ref{sec:IMDP_abstraction}).

The system ${\mathbf{S}}$ can equivalently be described using a stochastic kernel ${\mathbf{t}} \colon {\mathcal{X}} \times {\mathcal{U}} \rightarrow \Delta({\mathcal{X}}, \mathcal{B}({\mathcal{X}}))$~\cite[Proposition~11.6]{kallenberg2006foundations}.
For all $x_k \in {\mathcal{X}}$, $u_k \in {\mathcal{U}}$, and $A \in \mathcal{B}({\mathcal{X}})$, this kernel is defined as
$
    {\mathbf{t}}(A \mid x_k, u_k) 
        = \mathbb{P}\left( w_k \in {\mathcal{W}} : f(x_k, u_k, w_k) \in A \right).
$
Thus, ${\mathbf{t}}(A \mid x_k, u_k) \in [0,1]$ is the probability that $x_{k+1} \in A$, conditioned on the state $x_k$ and input $u_k$.

\begin{example} \label{ex:dubins}
    Throughout the paper, we use the example of a 3D-state Dubins car whose dynamics are defined as
    \begin{equation}
    \begin{split}
        x_{k+1} &= x_k + \tau u_k^{[2]}\cos{\theta_k}, \\ 
        y_{k+1} &= y_k + \tau u_k^{[2]}\sin{\theta_k}, \\ 
        \theta_{k+1} &= \text{\textup{wrap}}\big[\theta_k + \tau \alpha  u_k^{[1]} + w_k\big],
    \end{split}
    \end{equation}
    where $x_k \in [-10,10]$, $y_k \in [-10,10]$, and $\theta_k \in [-\pi,\pi]$ are the (x,y)-coordinates and the steering angle, respectively, and the two inputs are the change in steering angle $u_k^{[1]}\in[-\pi/2,\pi/2]$ and the driving speed $u_k^{[2]}\in[-3,3]$.
    The change in steering angle is noisy, modeled by the stochastic disturbances $w_k \sim \mathcal{N}(0, 0.01)$. 
    The parameters are set as $\tau = 1$ and $\alpha = 0.85$.
    The function `$\text{\textup{wrap}}$' brings the angle update $\theta_{k+1}$ within the range $[-\pi,\pi]$. 
\end{example}

\textbf{Policies\,\,}
The actions in the system ${\mathbf{S}}$ are selected by a memoryless policy ${\pi} \colon {\mathcal{X}} \to {\mathcal{U}}$, which is a universally measurable function.
The set of all such policies is denoted by ${\Pi_\mathbf{S}} = \{ {\mathcal{X}} \to {\mathcal{U}} \}$.
For a fixed policy ${\pi}$, the sequence of states $(x_0, x_1, \ldots)$ is obtained as $x_0 \sim {\mu_{x_0}}$ and $x_{k+1} \sim {\mathbf{t}}(\cdot \mid x_k, {\pi}(x_k))$ for all $k \in {\mathbb{N}}_0$.

\textbf{Specifications\,\,}
To define control tasks (called \emph{specifications}), we equip the system ${\mathbf{S}}$ with a universally measurable \emph{labeling function} ${h_{\mathbf{S}}} \colon {\mathcal{X}} \to 2^{\mathcal{Y}}$ over a finite set of labels ${\mathcal{Y}}$.
Intuitively, the labeling function \emph{tags} the state space with regions of interest.
In this paper, we focus on (infinite-horizon) \emph{reach-avoid specifications}, which require the system to eventually reach the goal states ${\mathcal{X}}_G \subset {\mathcal{X}}$ while avoiding unsafe states ${\mathcal{X}}_U \subset {\mathcal{X}}$ until then.
Such a specification uses the labels ${\mathcal{Y}} = \{\mathsf{goal}, \mathsf{safe}\}$ and labeling function ${h_{\mathbf{S}}}$ defined for all $x \in {\mathcal{X}}$ as
\begin{align*}
    \mathsf{goal} \in {h_{\mathbf{S}}}(x) &\iff x \in {\mathcal{X}}_G, \enskip \\
    \mathsf{safe} \in {h_{\mathbf{S}}}(x) &\iff x \in {\mathcal{X}} \setminus {\mathcal{X}}_U.
\end{align*}
The reach-avoid specification is then identified by its set of satisfying output traces $\mathbf{Y}$, defined as
\begin{equation}
\begin{split}
    \label{eq:RA_spec}
    \mathbf{Y} \coloneqq   \big\{  
        \{{h_{\mathbf{S}}}(x_k)\}_{k \in {\mathbb{N}}_0} 
        : 
        {}&{} \exists k \in {\mathbb{N}}. \,  \mathsf{goal} \in {h_{\mathbf{S}}}(x_k) \, \wedge 
        \\ 
        {}&{} \forall k' \leq k. \, \mathsf{safe} \in {h_{\mathbf{S}}}(x_{k'}) 
    \big\}.
\end{split}
\end{equation}
Intuitively, the set $\mathbf{Y}$ thus contains all the output traces that satisfy the control task.

\begin{remark}
\label{remark:DFA}
More general specifications expressed in, e.g., (syntactically co-safe) linear temporal logic~(LTL), require policies to have memory~\cite{DBLP:conf/focs/Pnueli77,belta2017formal}.
Our techniques are compatible with these specifications via the standard approach \cite{LSAZ21}: express the specification as an automaton with states~${\mathcal{Q}}$, construct the product of system~${\mathbf{S}}$ and the automaton, and compute a policy ${\pi} \colon {\mathcal{X}} \times {\mathcal{Q}} \to {\mathcal{U}}$ on this product state space.
A special case is the reach-avoid specification over a finite horizon of ${\ensuremath{K}} \in {\mathbb{N}}$ steps, in which case the states ${\mathcal{Q}} = \{0,1,\ldots,{\ensuremath{K}}\}$ encode precisely the time steps up to the horizon, and the policy is \emph{Markovian}, i.e., has the form ${\pi} \colon {\mathcal{X}} \times \{0,1,\ldots,{\ensuremath{K}}\} \to {\mathcal{U}}$~\cite{Bertsekas.Shreve78}.
For simplicity, we restrict ourselves to specifications that do not need this product construction and focus on infinite-horizon reach-avoid specifications instead.
\end{remark}

\begin{figure}[t]
    \centering
    \begin{subfigure}[t]{0.549\linewidth}
        \includegraphics[width=\linewidth]{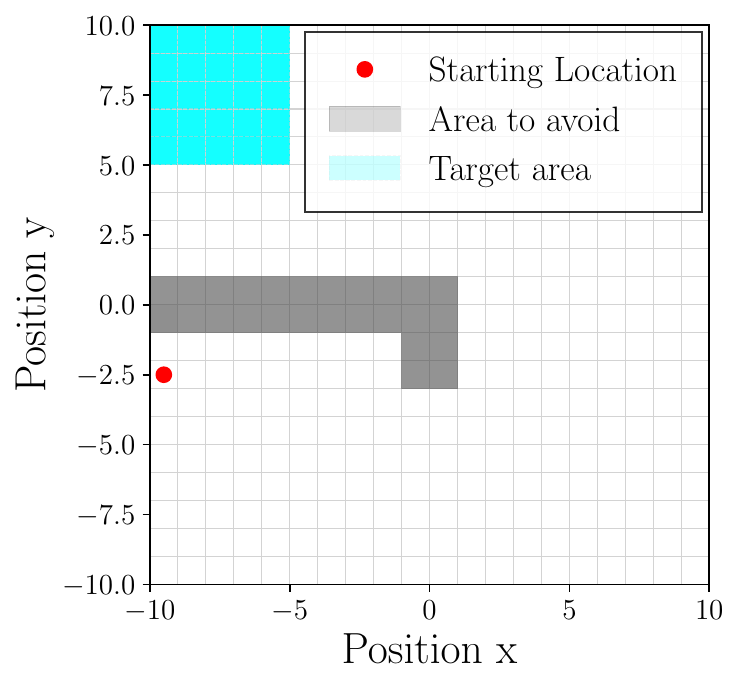}
        \caption{State-space partition}
        \label{fig:abstraction-state}
    \end{subfigure} % 
    \hfill
    \begin{subfigure}[t]{.4\linewidth}
        \includegraphics[width=0.785\linewidth]{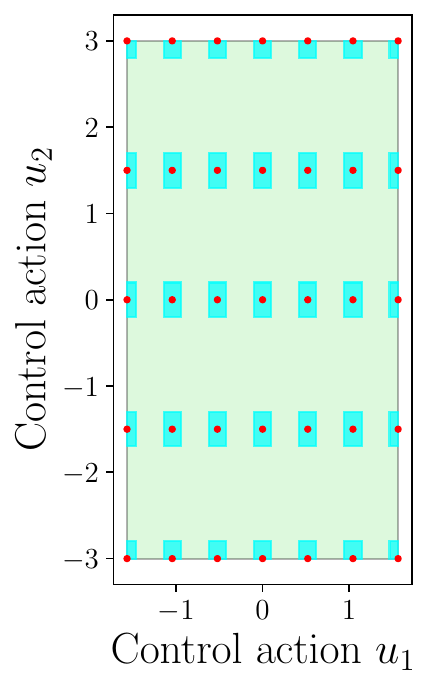}
        \caption{Input discretization}
        \label{fig:abstraction-input}
    \end{subfigure} 
    \caption{Our abstraction is based on a partition of the state space ${\mathcal{X}}$ (left; showing a reach-avoid problem with goal states in cyan and avoid states in grey, and each abstract action $a_i$ corresponds to an $L_p$-ball $B(u_i,\epsilon_i)$ in the input space ${\mathcal{U}}$ (right; with $L_p$-balls in cyan and their centers in red). 
    }
    \label{fig:abstraction}%
\end{figure}

\begin{example}[continues=ex:dubins]
    Suppose that the Dubins car needs to reach ${\mathcal{X}}_G = [-10,-5] \times [5,10] \times [-\pi,\pi]$ while avoiding the dark states ${\mathcal{X}}_U$ shown in Fig.~\ref{fig:abstraction} and while not leaving ${\mathcal{X}}$.
    This reach-avoid specification is identified by the set of output traces $\mathbf{Y}$ defined in \eqref{eq:RA_spec}.
\end{example}

\textbf{Closed-loop system\,\,}
Fixing a policy ${\pi}$ creates a stochastic process over paths $\{ x_k \}_{k \in {\mathbb{N}}_0}$ and thus also over output traces $\{ y_k \}_{k \in {\mathbb{N}}_0} = \{ {h_{\mathbf{S}}}(x_k) \}_{k \in {\mathbb{N}}_0}$.
This \emph{closed-loop system} is defined on the sample space $\Omega \coloneqq \prod_{k=1}^\infty (2^{\mathcal{Y}})$ endowed with its product topology $\mathcal{B}(\Omega)$ and a probability measure ${\mathbb{P}}^{{\pi}}_{{\mathbf{S}}}$ over output traces $(y_0, y_1, \ldots)$ uniquely generated by the transition kernel ${\mathbf{t}}$~\cite[Proposition~7.45]{Bertsekas.Shreve78}.
Intuitively, ${\mathbb{P}}^{{\pi}}_{{\mathbf{S}}}(\mathbf{Y})$ \emph{measures} the probability that the closed-loop system generates an output trace $(y_0, y_1, \ldots)$ contained in $\mathbf{Y} \subseteq (2^{\mathcal{Y}})^{{\mathbb{N}}_0}$.

\subsection{Problem statement}
Our goal is to compute a policy ${\pi}$ for the system ${\mathbf{S}}$ that (i)~minimizes a desired cost function, and (ii)~guarantees that a given specification is satisfied with at least a desired probability of $\lambda \in [0,1]$.
This cost function is defined as a real-valued \emph{cost function} $J \colon ({\mathcal{X}} \times {\mathcal{U}})^{{\mathbb{N}}_0} \to {\mathbb{R}}$ over paths and can model various performance indicators, such as total control effort, and even supports non-Markovian objectives.
We write ${\mathbb{E}}_{\mathbf{S}}^{\pi}[J(x_0, u_0, x_1, u_1, \cdots)]$ for the expectation over this cost function in the system ${\mathbf{S}}$ for the fixed policy ${\pi}$, i.e.,\ w.r.t.\ the probability ${\mathbb{P}}^{{\pi}}_{{\mathbf{S}}}({\mathbf{Y}})$.
Then, the problem we consider is stated as follows:

\begin{problem}
    \label{problem}
    Consider a system ${\mathbf{S}}$ as in \eqref{eq:dynamics}, a reach-avoid specification $\mathbf{Y}$ as in \eqref{eq:RA_spec}, a threshold $\lambda \in [0,1]$, and a cost function $J \colon ({\mathcal{X}} \times {\mathcal{U}})^{{\mathbb{N}}_0} \to {\mathbb{R}}$.
    Compute a policy ${\pi}$ such that ${\mathbb{P}}^{{\pi}}_{{\mathbf{S}}}({\mathbf{Y}}) \geq \lambda$ and that minimizes ${\mathbb{E}}_{\mathbf{S}}^{\pi}[J(x_0, u_0, x_1, u_1, \cdots)]$, i.e., a policy that solves the optimization problem:
    \begin{equation}
    \begin{split}
        \label{eq:policy-opt}
        \min_{\pi\in{\Pi_\mathbf{S}}} \,\,\, & {\mathbb{E}}_{\mathbf{S}}^{\pi}[J(x_0, u_0, x_1, u_1, \cdots)]
        \\
        s.t. \,\,\, & {\mathbb{P}}^{{\pi}}_{{\mathbf{S}}}({\mathbf{Y}}) \geq \lambda.
    \end{split}
    \end{equation}
\end{problem}

\begin{example}
    Suppose we want to synthesize a policy that satisfies the reach-avoid specification ${\mathcal{Y}}$ in~\eqref{eq:RA_spec} with probability of at least $\lambda = 0.9$.
    In addition, suppose we want to minimize the expected total control effort 
    modeled by the cost function ${\mathbb{E}}[J] = {\mathbb{E}}[\sum_{k \in {\mathbb{N}}_{0}} \|u_k\|^2_2]$.
    Standard abstraction techniques optimize solely for this satisfaction probability but not for the cost function $J$.
    Our goal is, instead, to additionally minimize the expected cost $J$.
    As a result, the satisfaction probability ${\mathbb{P}}^{{\pi}}_{{\mathbf{S}}}({\mathbf{Y}})$ may be slightly lower, but it cannot violate the threshold of $\lambda$.
\end{example}

Solving Prob.~\ref{problem} exactly amounts to solving a stochastic and nonconvex optimization program that is generally intractable~\cite{Bertsekas.Shreve78}.
Thus, we will treat ${\mathbb{P}}^{{\pi}}_{{\mathbf{S}}}({\mathbf{Y}}) \geq \lambda$ as a \emph{hard constraint} and minimizing ${\mathbb{E}}_{\mathbf{S}}^{\pi}[J(x_0, u_0, x_1, u_1, \cdots)]$ as a \emph{soft requirement}.
That is, we seek a policy that satisfies ${\mathbf{Y}}$ with probability $\geq\lambda$, and that yields a low (but possibly suboptimal) expected cost ${\mathbb{E}}_{\mathbf{S}}^{\pi}[J(x_0, u_0, x_1, u_1, \cdots)]$.

\textbf{Overview\,\,}
Our approach to solving Prob.~\ref{problem} is based on a finite-state abstraction in the form of an IMDP (see Def.~\ref{def:IMDP} below).
Existing abstractions are predominantly based on a partition of the state space and a gridding of the action space~\cite{DBLP:journals/tac/LahijanianAB15,Badings2025CDC,DBLP:conf/hybrid/MathiesenHL25,DBLP:journals/tac/HaesaertS21}.
We, by contrast, define in Sec.~\ref{sec:IMDP_abstraction} a novel and more general abstraction: the discrete states are still based on a partition, but each abstract action corresponds to a \emph{set of inputs} (instead of a single input), 
leading to the novel notion of a \emph{set-valued interface function} (defined in Sec.~\ref{sec:Interface}).
Doing so, we offload part of the control to an online control algorithm (for which we use MPC; see Sec.~\ref{sec:MPC}) that further refines the policy by selecting precise inputs from these sets.

\section{Interval MDPs}
\label{sec:IMDPs}
We use interval Markov decision processes (IMDPs)~\cite{suilen_RobustMarkovDecision_2025} to represent finite-state abstractions of the system ${\mathbf{S}}$.

\begin{definition}[IMDP]
    \label{def:IMDP}
    An \emph{interval MDP} (IMDP) ${\mathbf{M}}$ is a tuple ${({\mathcal{S}},{\mu_{s_0}},{\mathcal{A}},{\check{P}},{\hat{P}},{\mathcal{Y}},{h_{\mathbf{M}}})}$, where
    \begin{itemize}
        \item ${\mathcal{S}}$ is a finite set of states,
        \item ${\mu_{s_0}} \in \Delta(S)$ is the initial state distribution,
        \item ${\mathcal{A}}$ is a finite set of actions, and we write ${\mathcal{A}}(s) \subseteq {\mathcal{A}}$ for the actions enabled in state $s \in S$,
        \item ${\check{P}}, {\hat{P}} \colon {\mathcal{S}} \times {\mathcal{A}} \times {\mathcal{S}} \rightharpoonup [0,1]$ assign\footnote{The functions ${\check{P}}$, ${\hat{P}}$, and ${\mathcal{P}}$ are partial maps, written as $\rightharpoonup$, to model that not all actions may be enabled in every state.} a lower and upper bound to each transition probability, respectively, which define the uncertain transition function ${\mathcal{P}} \colon {\mathcal{S}} \times {\mathcal{A}} \rightharpoonup 2^{\Delta({\mathcal{S}})}$ for all $s \in {\mathcal{S}}, a \in {\mathcal{A}}(s)$ as
        \begin{align*}
            {\mathcal{P}}(s,a) = \big\{ {}&{} \mu \in \Delta({\mathcal{S}}) : 
            \forall s' \in {\mathcal{S}}, \\ 
            {}&{} \, \mu(s') \in [{\check{P}}(s,a,s'), {\hat{P}}(s,a,s')]
            \big\},
        \end{align*}
        \item ${\mathcal{Y}}$ is a finite set of labels, and
        \item ${h_{\mathbf{M}}} \colon {\mathcal{S}} \to 2^{\mathcal{Y}}$ is a labeling function over ${\mathcal{Y}}$.
     \end{itemize}
\end{definition}

We call $[{\check{P}}(s,a,s'), {\hat{P}}(s,a,s')] \subseteq [0,1]$ the \emph{probability interval} for the transition $(s,a,s')$.
Analogous to the system ${\mathbf{S}}$, the actions in an IMDP are selected by a memoryless policy ${\sigma} \colon {\mathcal{S}} \to {\mathcal{A}}$.\footnote{For clarity, we write policies for the system ${\mathbf{S}}$ as ${\pi}$ and policies for the IMDP ${\mathbf{M}}$ as ${\sigma}$.}
The set of all policies for ${\mathbf{M}}$ is written as ${\Pi_\mathbf{M}}$.

Intuitively, an IMDP defines a set of MDPs that differ only in their transition probabilities.\footnote{Interpreting an IMDP as a set of MDPs is known as the \emph{static} uncertainty model, as opposed to the \emph{dynamic} model where different probabilities can be chosen every time the same state-action pair is reached~\cite{suilen_RobustMarkovDecision_2025}. For IMDPs, both interpretations coincide for the optimality criterion in \eqref{eq:optimal_policy}~\cite{DBLP:journals/mor/Iyengar05}.
}
We overload notation and write $P \in {\mathcal{P}}$ for fixing a distribution $P(s,a) \in {\mathcal{P}}(s,a)$ for all $s \in {\mathcal{S}}$, $a \in {\mathcal{A}}(s)$.
Thus, fixing a policy ${\sigma} \in {\Pi_\mathbf{M}}$ and transition function $P \in {\mathcal{P}}$ for ${\mathbf{M}}$ yields a Markov chain with (standard) probability measure ${\mathbb{P}}_{\mathbf{M}}^{{\sigma},P}$ over paths~\cite{BaierKatoen08}.
Analogous to the system ${\mathbf{S}}$, a reach-avoid specification ${\mathbf{Y}}'$ for ${\mathbf{M}}$ is identified by its set of satisfying output traces, such that ${\mathbf{Y}}' \subseteq \prod_{k=1}^\infty (2^{\mathcal{Y}})$.
The probability that ${\mathbf{M}}$ with policy ${\sigma}$ and transition function~$P$ satisfies ${\mathbf{Y}}'$ is written as ${\mathbb{P}}_{\mathbf{M}}^{{\sigma},P}({\mathbf{Y}}')$.
An \emph{optimal (robust) policy} ${\sigma}^\star \in {\Pi_\mathbf{M}}$ maximizes this probability under the worst-case adversary:\footnote{Conversely, we may define variants of robust policies for IMDPs where the $\max$ and/or $\min$ operators are inverted.}
\begin{equation}
    \label{eq:optimal_policy}
    {\sigma}^\star \in \argmax_{{\sigma} \in {\Pi_\mathbf{M}}} \, \min_{P \in {\mathcal{P}}} {\mathbb{P}}_{\mathbf{M}}^{{\sigma},P}({\mathbf{Y}}').
\end{equation}
As the IMDP has finitely many states and actions, optimal robust policies can be computed efficiently using \emph{robust value iteration}~\cite{DBLP:conf/cdc/WolffTM12,DBLP:journals/tac/HaesaertS21}, implemented in probabilistic model checkers such as PRISM~\cite{DBLP:conf/cav/KwiatkowskaNP11} and Storm~\cite{DBLP:journals/sttt/HenselJKQV22}.

\section{Set-Valued Interface Functions}
\label{sec:Interface}
At the core of an abstraction of system ${\mathbf{S}}$ into an IMDP ${\mathbf{M}}$ is a (measurable) \emph{relation} ${\mathcal{R}} \subseteq {\mathcal{X}} \times {\mathcal{S}}$ between the states of both models.
In this work, we only consider relations such that $|{\mathcal{R}}(x)| = 1$ for all $x \in {\mathcal{X}}$, i.e., ${\mathcal{R}}$ represents a \emph{partition} of ${\mathcal{X}}$.\footnote{For more general abstractions based on, for example, a cover of the state space, see, e.g.,~\cite{DBLP:journals/tac/HaesaertS21}. However, these abstractions lead to more involved refinement strategies.}
We overload notation and write such a relation as a function ${\mathcal{R}} \colon {\mathcal{X}} \to {\mathcal{S}}$, whose preimage is defined as ${\mathcal{R}}^{-1}(s) = \{ x \in {\mathcal{X}} \mid {\mathcal{R}}(x) = s \}$.

Next, we define the notion of an \emph{interface function}, commonly used in abstraction-based control~\cite{DBLP:journals/automatica/GirardP09,DBLP:journals/tac/HaesaertS21,DBLP:conf/hybrid/HuijgevoortSSH23,Badings2025CDC}.

\begin{definition}[Interface]
    \label{def:interface}
    An \emph{interface function} for the system ${\mathbf{S}}$, the IMDP ${\mathbf{M}}$, and the relation ${\mathcal{R}} \colon {\mathcal{X}} \to {\mathcal{S}}$ is a function ${\mathcal{F}} \colon {\mathcal{X}} \times {\mathcal{A}} \to {\mathcal{U}}$ that maps every $x \in {\mathcal{X}}$ and $a \in {\mathcal{A}}$ to an input $u \in {\mathcal{U}}$ for ${\mathbf{S}}$.
\end{definition}

An interface function refines every state $x \in {\mathcal{X}}$ and abstract action $a \in {\mathcal{A}}$ into a \emph{single} input $u \in {\mathcal{U}}$ for the system ${\mathbf{S}}$.
As a result, the interface completely determinizes the stochastic system ${\mathbf{S}}$, without leaving any room to optimize for the cost function $J$ in Prob.~\ref{problem}.
Therefore, we propose the following novel definition that generalizes the interface to be a map from states to \emph{sets of inputs}.

\begin{definition}[Set-valued interface]
    \label{def:interface_set}
    A \emph{set-valued interface function} for the system ${\mathbf{S}}$, the IMDP ${\mathbf{M}}$, and the relation ${\mathcal{R}} \colon {\mathcal{X}} \to {\mathcal{S}}$ is a function ${\mathcal{F}_\textnormal{set}} \colon {\mathcal{X}} \times {\mathcal{A}} \to 2^{\mathcal{U}}$ that maps every $x \in {\mathcal{X}}$ and $a \in {\mathcal{A}}$ to a set of inputs for ${\mathbf{S}}$.
\end{definition}

A (standard) interface function ${\mathcal{F}}$ is a special case of a set-valued interface function ${\mathcal{F}_\textnormal{set}}$ with singleton inputs.

The next notion is that of a lifting for a relation ${\mathcal{R}} \subseteq {\mathcal{X}} \times {\mathcal{S}}$, as defined in~\cite{DBLP:journals/siamco/HaesaertSA17}, that lifts a relation on two state spaces to distributions over these state spaces.

\begin{definition}[Lifting~\cite{DBLP:journals/siamco/HaesaertSA17}]
    \label{def:lifting}
    Let ${\mathcal{R}} \subseteq {\mathcal{X}} \times {\mathcal{S}}$ be a relation for $({\mathcal{X}}, \mathcal{B}({\mathcal{X}}))$ and $({\mathcal{S}}, \mathcal{B}({\mathcal{S}}))$.
    The relation ${{\mathcal{R}}^\mathcal{P}} \subseteq \Delta({\mathcal{X}}, \mathcal{B}({\mathcal{X}})) \times \Delta({\mathcal{S}}, \mathcal{B}({\mathcal{S}}))$ is called a \emph{lifting of relation ${\mathcal{R}}$} if $(\mu, \nu) \in {{\mathcal{R}}^\mathcal{P}}$ holds for all $\mu \in \Delta({\mathcal{X}}, \mathcal{B}({\mathcal{X}}))$ and $\nu \in \Delta({\mathcal{S}}, \mathcal{B}({\mathcal{S}}))$ for which there exists a probability space $({\mathcal{X}} \times {\mathcal{S}}, \mathcal{B}({\mathcal{X}} \times {\mathcal{S}}), \mathbb{W})$ satisfying:
    \begin{enumerate}
        \item for all $A \in \mathcal{B}({\mathcal{X}})$ it holds that $\mathbb{W}(A, {\mathcal{S}}) = \mu(A)$,
        \item for all $B \in \mathcal{B}({\mathcal{S}})$ it holds that $\mathbb{W}({\mathcal{X}}, B) = \nu(B)$,
        \item all probability mass is on $\mathcal{R}$, i.e., $\mathbb{W}({\mathcal{R}}) = 1$.
    \end{enumerate}
\end{definition}

Based on the set-valued interface and lifted relation, we now define our novel relation between the system ${\mathbf{S}}$ and the IMDP ${\mathbf{M}}$ that is fundamental to solving Prob.~\ref{problem}.

\begin{definition}[Probabilistic alternating simulation]
    \label{def:PASR}
    Consider a system ${\mathbf{S}}$ as in \eqref{eq:dynamics} and an IMDP ${\mathbf{M}} = {({\mathcal{S}},{\mu_{s_0}},{\mathcal{A}},{\check{P}},{\hat{P}},{\mathcal{Y}},{h_{\mathbf{M}}})}$.
    If there exist
    \begin{itemize}
        \item a relation ${\mathcal{R}} \subseteq {\mathcal{X}} \times {\mathcal{A}}$ with $|{\mathcal{R}}(x)| = 1$ for all $x \in {\mathcal{X}}$,
        \item a lifting ${{\mathcal{R}}^\mathcal{P}}$ of the relation ${\mathcal{R}}$, and
        \item a set-valued interface function ${\mathcal{F}_\textnormal{set}} \colon {\mathcal{X}} \times {\mathcal{A}} \to 2^{\mathcal{U}}$,
    \end{itemize}
    such that the following conditions are satisfied:
    \begin{enumerate}
        \item initial distributions are related: $({\mu_{x_0}}, {\mu_{s_0}}) \in {{\mathcal{R}}^\mathcal{P}}$,
        \item for all $(x,s) \in {\mathcal{R}}$, labels coincide: ${h_{\mathbf{S}}}(x) = {h_{\mathbf{M}}}(s)$,
        \item for all $(x,s) \in {\mathcal{R}}$, $a \in {\mathcal{A}}$, and $u \in {\mathcal{F}_\textnormal{set}}(x, a)$, there exists $\nu \in {\mathcal{P}}(s,a)$ such that 
        $
            \big( {\mathbf{t}}(\cdot \mid x, u), \, \nu \big) \in {{\mathcal{R}}^\mathcal{P}},
        $
    \end{enumerate}
    then the relation ${\mathcal{R}}$ is a \emph{probabilistic alternating simulation relation} (PASR) from ${\mathbf{S}}$ to ${\mathbf{M}}$, denoted by ${\mathbf{S}} \preceq_{\textnormal{alt}} {\mathbf{M}}$.
\end{definition}

The PASR in Def.~\ref{def:PASR} is a variant of \cite[Def.~7]{Badings2025CDC}, tailored to one system with a precise stochastic kernel (the system ${\mathbf{S}}$) and one with an uncertain stochastic kernel (the IMDP ${\mathbf{M}}$).
As a novel extension, we generalize the interface function to sets of inputs, requiring an additional quantification over all $u \in {\mathcal{F}_\textnormal{set}}(x, a)$ in condition~$(c)$.

Intuitively, the third condition asserts that for all pairs of related states $(x,s) \in {\mathcal{R}}$, every IMDP action $a \in {\mathcal{A}}$, and for every possible refined control input $u \in {\mathcal{F}_\textnormal{set}}(x,a)$, the distribution over next states ${\mathcal{X}}$ in the system ${\mathbf{S}}$ is related to the distribution over states ${\mathcal{S}}$ in the IMDP ${\mathbf{M}}$.
As the main result of this section, we show that the satisfaction probability under \emph{any} policy $\pi$ that can be obtained through the interface ${\mathcal{F}_\textnormal{set}}$ is lower (resp. upper) bounded by the minimum (resp. maximum) satisfaction probability on the IMDP:

\begin{theorem}[Policy refinement]
    \label{thm:PASR_synthesis}
    Let ${\mathbf{S}} \preceq_{\textnormal{alt}} {\mathbf{M}}$.
    Then for every IMDP policy ${\sigma} \colon {\mathcal{S}} \to {\mathcal{A}}$ and every specification ${\mathbf{Y}} \subseteq \prod_{k=1}^\infty (2^{\mathcal{Y}})$, it holds that
    \begin{equation}   
        \label{eq:PASR_synthesis}
        \min_{P \in {\mathcal{P}}} {\mathbb{P}}_{\mathbf{M}}^{{\sigma},P}({\mathbf{Y}})
        \leq
        {\mathbb{P}}^{{\pi}}_{{\mathbf{S}}}({\mathbf{Y}})
        \leq
        \max_{P \in {\mathcal{P}}} {\mathbb{P}}_{\mathbf{M}}^{{\sigma},P}({\mathbf{Y}}),
    \end{equation}
    where the policy ${\pi}$ is defined for all 
    $x \in {\mathcal{X}}$ as
    \begin{equation}
        \label{eq:PASR_policy}
        {\pi}(x) \in {\mathcal{F}_\textnormal{set}}(x, {\sigma}({\mathcal{R}}(x))).
    \end{equation}
\end{theorem}

The proof of Thm.~\ref{thm:PASR_synthesis} is provided in App.~\ref{appendix:PASR_synthesis}.
Intuitively, Thm.~\ref{thm:PASR_synthesis} asserts that the existence of a PASR implies that the satisfaction probability ${\mathbb{P}}^{{\pi}}_{{\mathbf{S}}}({\mathbf{Y}})$ is bounded by the min/max satisfaction probability on the IMDP, as long as the policy ${\pi}$ for ${\mathbf{S}}$ is selected based on the set-valued interface function ${\mathcal{F}_\textnormal{set}}$.
To solve Prob.~\ref{problem}, we thus simply seek an IMDP policy ${\sigma}$ such that $\min_{P \in {\mathcal{P}}} {\mathbb{P}}_{\mathbf{M}}^{{\sigma},P}({\mathbf{Y}}) \geq \lambda$, meaning we can freely optimize the policy ${\pi}$ within ${\mathcal{F}_\textnormal{set}}$ to optimize the expected cost ${\mathbb{E}}_{\mathbf{S}}^{\pi}[J(x_0, u_0, x_1, u_1, \cdots)]$ while preserving ${\mathbb{P}}^{{\pi}}_{{\mathbf{S}}}({\mathbf{Y}}) \geq \lambda$.

\section{IMDP Abstractions with Set-Valued Actions}
\label{sec:IMDP_abstraction}
In this section, we present a model-based IMDP abstraction procedure that induces the PASR defined in Def.~\ref{def:PASR}.
Our abstraction procedure is a variant of~\cite{DBLP:journals/tac/LahijanianAB15,Badings2025CDC}, adapted to the set-valued interface function in Def.~\ref{def:interface_set}.
In what follows, we define the IMDP's states ${\mathcal{S}}$, actions ${\mathcal{A}}$, transition probabilities ${\check{P}}, {\hat{P}}$, and labeling function ${h_{\mathbf{M}}}$.

\textbf{States and labels\,\,}
We define the states by a partition of the state space ${\mathcal{X}}$ into $L \in {\mathbb{N}}$ non-overlapping and non-empty regions $\left\{{\mathcal{V}}_1,\ldots,{\mathcal{V}}_{L}\right\}$.
We define one IMDP state for each of the $L$ regions, such that ${\mathcal{S}} \coloneqq \{s_1,\ldots,s_L\}$.
A common choice is to define each region ${\mathcal{V}}_i$, $i=1,\ldots,{L-1}$, as a convex polytope and define ${\mathcal{V}}_L = {\mathcal{X}} \setminus \cup_{i=1}^{L-1} {\mathcal{V}}_i$.\footnote{\label{fn:partition}Technically, states $x$ on the boundary between regions are related to multiple IMDP states. However, Assumption~\ref{assumption:disturbance} means the probability of reaching such a state is zero, so this technicality does not affect the correctness of the abstraction.}
The partition induces a relation ${\mathcal{R}} \subseteq {\mathcal{X}} \times {\mathcal{S}}$ such that $(x,s_i) \in {\mathcal{R}} \iff x \in {\mathcal{V}}_i$.
For simplicity, we assume the partition is \emph{label-preserving}, allowing us to satisfy condition (2) in Def.~\ref{def:PASR}:

\begin{assumption}[Label-preserving]
    \label{assumption:labels}
    For all $x,x' \in {\mathcal{X}}$ such that ${\mathcal{R}}(x) = {\mathcal{R}}(x')$, it holds that ${h_{\mathbf{S}}}(x) = {h_{\mathbf{S}}}(x')$.
\end{assumption}

Observe that, for a label-preserving partition, the abstract labeling function ${h_{\mathbf{M}}}$ is trivially defined for all $s \in {\mathcal{S}}$ as ${h_{\mathbf{M}}}(s) \coloneqq {h_{\mathbf{S}}}(x)$, for any arbitrary $x \in {{\mathcal{R}}^{{-1}}}(s)$.

\begin{remark}
    \label{remark:label_preserving}
    Constructing a label-preserving polyhedral partition can be challenging (or even impossible, e.g., a circular goal region cannot be represented using finitely many polytopes).
    At the cost of weaker bounds on the satisfaction probability in Thm.~\ref{thm:PASR_synthesis}, one can weaken Assumption~\ref{assumption:labels} by suitably over/underapproximating the labels, or by defining a metric on the output space as in~\cite{DBLP:journals/siamco/HaesaertSA17}.
\end{remark}

\textbf{Actions\,\,}
In a standard abstraction, each abstract action $a_i \in {\mathcal{A}}$ corresponds to a discrete input $u_i \in {\mathcal{U}}$.
To accommodate the set-valued interface function, we instead associate each $a_i$ to a \emph{set of inputs}, defined as an $L_p$-ball, as also visualized in Fig.~\ref{fig:abstraction}.

\begin{definition}
    \label{def:Lp_ball}
    The $L_p$-ball, $p\in\mathbb{N}$, of radius $\epsilon > 0$ centered at $c \in {\mathbb{R}}^n$ is defined as
    $
    {B}(c, \epsilon) = \{ c' \in \mathbb{R}^n : \|c' - c\|_p \le \epsilon \}.
    $
\end{definition}

We define the set of IMDP actions as ${\mathcal{A}} \coloneqq \{ a_1,\ldots,a_M \}$, where each $a_i \in {\mathcal{A}}$ is associated with an $L_p$-ball ${B}(u_i, \epsilon_i)$ of radius $\epsilon_i$ centered at $u_i \in {\mathcal{U}}$.
We define the set-valued interface function ${\mathcal{F}_\textnormal{set}} \colon {\mathcal{X}} \times {\mathcal{A}} \to 2^{\mathcal{U}}$ for all $x \in {\mathcal{X}}$, $a_i \in {\mathcal{A}}$ as ${\mathcal{F}_\textnormal{set}}(x, a_i) = {B}(u_i, \epsilon_i) \cap {\mathcal{U}}$.
We discuss a possible criterion for selecting the size $\epsilon$ of each ball in Sec.~\ref{subsec:criteria_balls}.

\begin{example}[continues=ex:dubins]
    For the Dubins car, suppose we partition the state space into $20 \times 20 \times 11$ cells, yielding the partition shown in Fig.~\ref{fig:abstraction} (left).
    Similarly, we define $7 \times 5$ abstract actions, associated with $L_\infty$-balls with uniformly gridded centers and a fixed radius $\epsilon = [0.2,0.4]$, as also shown in Fig.~\ref{fig:abstraction} (right).
    Note that in a conventional IMDP abstraction (e.g., as in~\cite{DBLP:journals/tac/LahijanianAB15,Badings2025CDC}), the $L_\infty$-balls around the centers would be absent.
\end{example}

\textbf{Transition probabilities\,\,}
For every triple $(s,a,s') \in {\mathcal{S}} \times {\mathcal{A}} \times {\mathcal{S}}$, we compute the probability interval $[{\check{P}}(s,a,s'), {\hat{P}}(s,a,s')]$ that defines the set of distributions ${\mathcal{P}}(s,a)$.
For the standard IMDP abstraction procedure~\cite{DBLP:journals/tac/LahijanianAB15,Badings2025CDC,DBLP:conf/hybrid/CauchiLLAKC19} (where each action $a_i$ corresponds to a singleton input $u_i$), this interval would be defined as
\[
\left[
    \min_{x \in {{\mathcal{R}}^{{-1}}}(s)} \! {\mathbf{t}}({{\mathcal{R}}^{{-1}}}(s') \mid x, u_i),
    \max_{x \in {{\mathcal{R}}^{{-1}}}(s)} \! {\mathbf{t}}({{\mathcal{R}}^{{-1}}}(s') \mid x, u_i)
\right],
\]
i.e., the interval is characterized by the min./max. probability of reaching the partition element ${{\mathcal{R}}^{{-1}}}(s')$ when starting from any state $x \in {{\mathcal{R}}^{{-1}}}(s)$ and executing the input $u_i$.

For our setting, we use the set-valued interface ${\mathcal{F}_\textnormal{set}}$ to additionally minimize and maximize over the ball ${B}(u_i, \epsilon_i)$. Thus, the probability bounds for transition $(s,a,s')$ are
\begin{align}
    \label{eq:IMDP:lb}
    {\check{P}}(s,a,s') &= \min_{\substack{x \in {{\mathcal{R}}^{{-1}}}(s) \\ u \in {\mathcal{F}_\textnormal{set}}(x,a)}} {\mathbf{t}}({{\mathcal{R}}^{{-1}}}(s') \mid x, u_i), \\
    \label{eq:IMDP:ub}
    {\hat{P}}(s,a,s') &= \max_{\substack{x \in {{\mathcal{R}}^{{-1}}}(s) \\ u \in {\mathcal{F}_\textnormal{set}}(x,a)}} {\mathbf{t}}({{\mathcal{R}}^{{-1}}}(s') \mid x, u_i).
\end{align}
For systems with additive (Gaussian) noise of diagonal covariance, these probabilities can be computed efficiently; see~\cite{DBLP:conf/hybrid/CauchiLLAKC19}.
For more general dynamics, sampling-based approaches can be used to derive \emph{probably approximately correct} (PAC) bounds on these probabilities~\cite{DBLP:journals/jair/BadingsRAPPSJ23}.

The full uncertain transition function ${\mathcal{P}} \colon {\mathcal{S}} \times {\mathcal{A}} \rightharpoonup 2^{\Delta({\mathcal{S}})}$ is obtained by computing the lower and upper bounds in \eqref{eq:IMDP:lb} and \eqref{eq:IMDP:ub} for every transition $(s,a,s')$.

\subsection{Policy synthesis and refinement}
Putting everything together, we thus construct an IMDP abstraction ${\mathbf{M}}$ with:
\begin{itemize}
    \item set of states ${\mathcal{S}} \coloneqq \{s_1,\ldots,s_L\}$;
    \item initial state distribution ${\mu_{s_0}}$ defined as ${\mu_{s_0}}(s) \coloneqq {\mu_{x_0}}({{\mathcal{R}}^{{-1}}}(s))$ for all $s \in {\mathcal{S}}$;
    \item set of actions ${\mathcal{A}} \coloneqq \{ a_1,\ldots,a_M \}$;
    \item probability bounds ${\check{P}}$ and ${\hat{P}}$ defined for all $(s,a,s')$ as in \eqref{eq:IMDP:lb} and \eqref{eq:IMDP:ub};
    \item set of labels ${\mathcal{Y}}$ the same as for the system ${\mathbf{S}}$;
    \item labeling function ${h_{\mathbf{M}}}$ defined for all $s \in {\mathcal{S}}$ as ${h_{\mathbf{M}}}(s) \coloneqq {h_{\mathbf{S}}}(x)$, for any arbitrary $x \in {{\mathcal{R}}^{{-1}}}(s)$.
\end{itemize}
The set-valued interface ${\mathcal{F}_\textnormal{set}} \colon {\mathcal{X}} \times {\mathcal{A}} \to 2^{\mathcal{U}}$ is defined as ${\mathcal{F}_\textnormal{set}}(x, a_i) = {B}(u_i, \epsilon_i) \cap {\mathcal{U}}$ for all $x \in {\mathcal{X}}$, $a_i \in {\mathcal{A}}$.
This IMDP is a probabilistic simulation of the original system, i.e., ${\mathbf{S}} \preceq_{\textnormal{alt}} {\mathbf{M}}$, as formalized by Thm.~\ref{thm:IMDP_correctness}.

\begin{theorem}
    \label{thm:IMDP_correctness}
    Let ${\mathbf{M}} = {({\mathcal{S}},{\mu_{s_0}},{\mathcal{A}},{\check{P}},{\hat{P}},{\mathcal{Y}},{h_{\mathbf{M}}})}$ be the IMDP abstraction for the system ${\mathbf{S}}$ obtained using the abstraction procedure above.
    Then it holds that~${\mathbf{S}} \preceq_{\textnormal{alt}} {\mathbf{M}}$.
\end{theorem}

We provide the proof of Thm.~\ref{thm:IMDP_correctness} in App.~\ref{appendix:IMDP_correctness}.
As a consequence of Thm.~\ref{thm:IMDP_correctness}, for any specification ${\mathbf{Y}}$, we can take any policy ${\sigma} \colon {\mathcal{S}} \to {\mathcal{A}}$ for the IMDP and use Thm.~\ref{thm:PASR_synthesis} to refine this policy ${\sigma}$ into a policy ${\pi}$ for the system ${\mathbf{S}}$ that satisfies the specification with probability at least $\min_{P \in {\mathcal{P}}} {\mathbb{P}}_{\mathbf{M}}^{{\sigma},P}({\mathbf{Y}})$ and at most $\max_{P \in {\mathcal{P}}} {\mathbb{P}}_{\mathbf{M}}^{{\sigma},P}({\mathbf{Y}})$.
\begin{corollary}
    \label{cor:refinement}
    Let ${\mathbf{M}} = {({\mathcal{S}},{\mu_{s_0}},{\mathcal{A}},{\check{P}},{\hat{P}},{\mathcal{Y}},{h_{\mathbf{M}}})}$ be the IMDP abstraction for the system ${\mathbf{S}}$ obtained using the abstraction procedure above, and fix an IMDP policy ${\sigma}$.
    Let $\pi_{\text{set}} \colon {\mathcal{X}} \to 2^{\mathcal{U}}$ be the permissive policy for the system ${\mathbf{S}}$ defined for all $x \in {\mathcal{X}}$ as
    \begin{equation}
        \label{eq:permissive_policy}
        \pi_{\text{set}}(x) = {\mathcal{F}_\textnormal{set}}(x, {\sigma}({\mathcal{R}}(x))).
    \end{equation}
    Then, for any policy $\pi \colon {\mathcal{X}} \to {\mathcal{U}}$ such that $\pi(x) \in \pi_{\text{set}}(x)$ for all $x \in {\mathcal{X}}$, it holds that
    \[
    \min_{P \in {\mathcal{P}}} {\mathbb{P}}_{\mathbf{M}}^{{\sigma},P}({\mathbf{Y}})
    \leq
    {\mathbb{P}}^{{\pi}}_{{\mathbf{S}}}({\mathbf{Y}})
    \leq
    \max_{P \in {\mathcal{P}}} {\mathbb{P}}_{\mathbf{M}}^{{\sigma},P}({\mathbf{Y}}).
    \]
\end{corollary}
Cor.~\ref{cor:refinement} follows directly from combining Thms.~\ref{thm:PASR_synthesis} and~\ref{thm:IMDP_correctness} and its proof is thus omitted.
If $\min_{P \in {\mathcal{P}}} {\mathbb{P}}_{\mathbf{M}}^{{\sigma},P}({\mathbf{Y}}) \geq \lambda$, then the refined policy ${\pi}$ is an admissible solution to Prob.~\ref{problem}.
In practice, we use robust value iteration to compute an optimal robust IMDP policy ${\sigma}^\star$ that \emph{maximizes} the probability of satisfying the specification on the IMDP under the \emph{worst-case} transition function, i.e., a policy ${\sigma}^\star$ satisfying \eqref{eq:optimal_policy}.
In our numerical experiments in Sec.~\ref{sec:Experiments}, we use robust value iteration implemented in the probabilistic model checker Storm~\cite{DBLP:journals/sttt/HenselJKQV22}.

\section{Abstraction-Driven Model Predictive Control\!}
\label{sec:MPC}

We now present the MPC controller that optimizes the cost function $J$.
We start with the overall control loop in Sec.~\ref{sec:MPC_control_loop} and present the MPC algorithm in Sec.~\ref{sec:MPC_architecture}.

\begin{algorithm}[t]
    \caption{Online feedback control loop\label{alg:online_control}}
    \begin{algorithmic}[1]
    \Require{IMDP policy ${\sigma}$, set-valued interface ${\mathcal{F}_\textnormal{set}}$}
    \State Let $x_0 \sim {\mu_{x_0}}$
    \For{$k = 0,1,\ldots$}
        \If{$\mathsf{goal} \in {h_{\mathbf{S}}}(x_k)$}
            \State \textbf{return} $\mathsf{SAT} = \mathsf{True}$
        \ElsIf{$\mathsf{safe} \notin {h_{\mathbf{S}}}(x_k)$}
            \State \textbf{return} $\mathsf{SAT} = \mathsf{False}$
        \EndIf
        \State Current IMDP state $s_k \gets {\mathcal{R}}(x_k)$
        \State IMDP action $a_k \gets {\sigma}(s_k)$
        \State Get target point $r_k \gets \boldsymbol{\textrm{R}}(x_k)$
        \State MPC input $u_k \gets \mathsf{MPC}(x_k, r_k) \in {\mathcal{F}_\textnormal{set}}(x_k, a_k)$
        \State Next state $x_{k+1} \sim {\mathbf{t}}(\cdot \mid x_k, u_k)$
    \EndFor
  \end{algorithmic}
\end{algorithm}

\subsection{Feedback control algorithm}
\label{sec:MPC_control_loop}
The online feedback control loop is described in Alg.~\ref{alg:online_control}.
At every time step $k \in {\mathbb{N}}$, we first determine the IMDP state $s_k \gets {\mathcal{R}}(x_k)$ associated with the measured state $x_k$ (Line 8) and retrieve the IMDP action $a_k$ (Line 9). Then we compute a target point $r_k$ (Line 10) for the state $x_k$ that we use as a reference point for the cost function. 
In practice, we choose this target point $r_k = \boldsymbol{\textrm{R}}(x_k)$ as the center of the partition element ${\mathcal{V}}_i$ that is part of the goal set ${\mathcal{X}}_G$, and that is closest to the current state $x_k$.
On Line 11, the MPC controller $\mathsf{MPC} \colon {\mathcal{X}} \times {\mathcal{X}} \to {\mathcal{U}}$, which we define concretely in the next subsection, selects \emph{any} input from the set of control inputs ${\mathcal{F}_\textnormal{set}}(x_k, a_k)$ associated with the action $a_k$. 
This MPC input $u_k = \mathsf{MPC}(x_k, r_k)$ is implemented on the system (Line 12).
The control loop terminates once a goal state (Lines 3-4) or an unsafe state is reached (Lines 5-6).

The policy described by Alg.~\ref{alg:online_control} satisfies Cor.~\ref{cor:refinement}, as the input $u_k \in {\mathcal{F}_\textnormal{set}}(x_k, {\sigma}({\mathcal{R}}(x_k)))$.
Thus, any MPC controller plugged into Line 11 of Alg.~\ref{alg:online_control} defines a feedback policy under which the closed-loop system satisfies the reach-avoid specification ${\mathbf{Y}}$ with probability at least $\min_{P \in {\mathcal{P}}} {\mathbb{P}}_{\mathbf{M}}^{{\sigma},P}({\mathbf{Y}})$.
If $\min_{P \in {\mathcal{P}}} {\mathbb{P}}_{\mathbf{M}}^{{\sigma},P}({\mathbf{Y}}) \geq \lambda$, then this policy is an admissible solution to Prob.~\ref{problem}.

\subsection{Model predictive control architecture}
\label{sec:MPC_architecture}
We now present a suitable MPC architecture for the control loop in Alg.~\ref{alg:online_control}.
Due to the nonlinear stochastic dynamics and piecewise constant form of the policy (over the state space partition), this MPC controller requires:
\begin{enumerate}
    \item a tractable yet accurate approximation of the stochastic dynamics $x_{k+1} = f(x_k,u_k,w_k)$, and
    \item a logic-driven structure to encode the $L_p$-balls of control inputs the MPC can select from.
\end{enumerate}
For point (a), we use piecewise affine approximations~\cite{sontag_NonlinearRegulationPiecewise_1981} of the nonlinear dynamics with a desired accuracy. 
For point (b), we resort to hybrid dynamical systems to represent the logic conditions linking measured and predicted states to respective cells in the partition of the abstraction and their associated $L_p$-balls. 
In the remainder of this section, we discuss the latter in more detail and present the complete MPC optimization problem.

\textbf{Logic-driven $L_p$-balls selection\,\,}
The logical constraints on the control inputs in the MPC are written as 
\begin{equation}
    x\in {\mathcal{R}}^{-1}(s_i) \Longrightarrow u\in {\mathcal{F}_\textnormal{set}}(x, \sigma(s_i)) = {B}(u_i, \epsilon).
\end{equation}
To embed these constraints into an MPC problem, we define $N_s = |S|$ as the number of states in the IMDP abstraction, and $N$ as the prediction horizon for the MPC problem. 
Then, we introduce $N_s \cdot (N+1)$ binary variables $\delta$ that satisfy the following conditions:
\begin{align}
    \label{eq:logical-conditions}
    & \delta_{k}^{i} = 1 \Longleftrightarrow x_k\in {\mathcal{R}}^{-1}(s_i),  \\
    & \sum_{i=1}^{N_s} \delta_k^{i} = 1 \quad \text{for}\,\, k=1,\ldots,N. \nonumber
\end{align}
For each region ${\mathcal{R}}^{-1}(s_i)$, we introduce the $n_x$-vectors $M_{s}^i$ and $m_{s}^i$ that represent the element-wise maximum and minimum boundaries of reach partition region, i.e.,
\begin{align} \label{eq:limits-state}
    & (M_{s}^i)_j = \max(({\mathcal{R}}^{-1}(s_i))_j),  \\
    & (m_{s}^i)_j = \min(({\mathcal{R}}^{-1}(s_i))_j), \nonumber
\end{align}
for $j=1,\ldots,n_x$ where $(\cdot)_j$ is the $j$-th component of the set. Following the procedure from~\cite{bemporad_ControlSystemsIntegrating_1999}, the logical conditions in \eqref{eq:logical-conditions} can be rewritten using the inequalities:
\begin{align}\label{eq:logical-inequalities}
    & m_{s}^i \delta_k^{i} \leq x_k \delta_k^{i}\leq M_{s}^i \delta_k^{i} \\ & \text{for} \,\, i = 1,\ldots,N_s; \,\, k=1,\ldots,N. \nonumber
\end{align}
These inequality constraints cannot be used directly into an optimization problem due to the nonlinearity introduced by the product $x \cdot \delta$.
Thus, we define $N_s \cdot (N+1)$ auxiliary variables $z_k^{i} = x_k \delta_k^{i}$ and rewrite \eqref{eq:logical-inequalities} as:
\begin{align} \label{eq:abstraction-consistency}
    & z_k^{i} \leq M_{s}^i \delta_k^{i} \\
    & z_k^{i} \geq m_{s}^i \delta_k^{i} \nonumber\\
    & z_k^{i} \leq x_k - m_{s}^i(1-\delta_k^{i}) \nonumber \\
    & z_k^{i} \geq x_k - M_{s}^i(1-\delta_k^{i}) \nonumber \\
    & \text{for} \,\, i = 1,\ldots,N_s; \,\, k=1,\ldots,N. \nonumber
\end{align}
These inequalities force $z_k^{i} = x_k^{i}$ if $x_k$ is in the region ${\mathcal{R}}^{-1}(s_i)$, i.e., $\delta_k^{i} = 1$, and make $z_k^{i} = 0$ elsewhere. In addition, \eqref{eq:abstraction-consistency} can only be satisfied if \eqref{eq:logical-conditions} holds, thus linking the state $x_k$ uniquely to an abstract state.

The next step is to link the control input to the policy. For this, we can use a linear combination of the $L_p$-balls. 
Using a definition similar to \eqref{eq:limits-state}, we introduce a number $N_s$ of $n_u$-vector boundaries $M_{a}^s$, $m_{a}^i$ for the $L_p$-balls:
\begin{align*}
    & (M_{a}^i)_j = \max({B}(u_i, \epsilon))_j), \quad (m_{a}^i)_j = \min(({B}(u_i, \epsilon))_j),
\end{align*}
and we use the same binary variables $\delta$ to write 
\begin{align} \label{eq:input-constraints}
    & \sum_{i = 1}^{N_s} m_{a}^i\delta_k^{i} \leq u_k \leq \sum_{i = 1}^{N_s} M_{a}^i\delta_k^{i} \\
    & \text{for} \,\, i = 1,\ldots,N_s; \,\, k=1,\ldots,N. \nonumber
\end{align}
As $x_k$ is contained in exactly one partition element, only one element of the sum in \eqref{eq:input-constraints} can be nonzero.
Thus, the control input can exclusively belong to the $L_p$-ball associated with the action chosen by the IMDP policy.

\textbf{MPC formulation\,\,}
We are now ready to present the complete MPC optimization problem, which minimizes a cost function $J(\tilde{\boldsymbol{x}}_k,\tilde{\boldsymbol{u}}_k,r_k)$ over the optimization window of $N \in {\mathbb{N}}$ steps, where the sequence state and input sequences are defined as $\tilde{\boldsymbol{x}}_k = [x_{1|k},\ldots,x_{N|k}]$ and $\tilde{\boldsymbol{u}}_k = [u_{0|k},\ldots,u_{N-1|k}]$, and $J$ is the sum over $N$ of the individual stage costs defined at each time step.
Here, we use the notation $j|k$ to represent the prediction step $j=1,\ldots,N$ given the measured state of the system at the time step $k$. Then, the optimization program associated to the MPC problem at time step $k$ is:
\begin{subequations}
\label{eq:MPC}
\begin{align}
    \min_{\substack{
            x_{1|k},\ldots,x_{N|k} \in{\mathcal{X}} \\
            u_{0|k},\ldots,u_{N-1|k}\in{\mathcal{U}} 
        }}
        \sum_{j=1}^N  J{}&{}(r_k-x_{j|k},u_{j-1|k}) \\
        \text{s.t.} \quad x_{k+1} &= f^{\text{MPC}}(x_k,u_k) \\
        x_{0|k} &= x_k \\
        z_{j|k}^{i} &\leq M_{s}^i \delta_{j|k}^{i} \\
        z_{j|k}^{i} &\geq m_{s}^i \delta_{j|k}^{i} \\
        z_{j|k}^{i} &\leq x_k - m_{s}^i(1-\delta_{j|k}^{i}) \\
        z_{j|k}^{i} &\geq x_k - M_{s}^i(1-\delta_{j|k}^{i}) \\
        & \enskip\text{for} \,\, i = 1,\ldots,N_s; \,\, j=0,\ldots,N \nonumber \\
        u_{j-1|k} &\leq \sum\nolimits_{i}^{N_s} M_{a}^i\delta_{j-1|k}^{i} \\
        u_{j-1|k} &\geq \sum\nolimits_{i = 0}^{N_s} m_{a}^i\delta_{j-1|k}^{i} \\
        & \enskip\text{for} \,\, j=1,\ldots,N \nonumber \\
        &\!\!\!\!\!\!\!\!\!\sum\nolimits_{i=1}^{N_s} \delta_{j|k}^{i} = 1 \\
        & \enskip\text{for} \,\, j=0,\ldots,N \nonumber 
\end{align}
\end{subequations}
In the experiments in Sec.~\ref{sec:Experiments}, we use \eqref{eq:MPC} with a standard quadratic stage cost $J(\tilde{\boldsymbol{x}}_k,\tilde{\boldsymbol{u}}_k,r_k) = \|r_k\cdot\boldsymbol{1}_{n_x} - \tilde{\boldsymbol{x}}_k\|_{Q} + \|\tilde{\boldsymbol{u}}_k\|_{R}$, where $\|\cdot\|_{Q}$ and $\|\cdot\|_{R}$ are the quadratic forms for positive definite matrices $Q$ and $R$.
This stage cost allows for a balanced optimization between the error from the reference point $r_k$ (Alg.~\ref{alg:online_control}, Line 10) and the control effort.

\section{Numerical Experiments}
\label{sec:Experiments}
We validate our framework on several benchmarks from the literature to answer two questions:
\begin{enumerate}
    \item[(Q1)] How does the size $\epsilon$ of the $L_p$-balls affect the lower bound $\lambda$ on the satisfaction probability?
    \item[(Q2)] Can our abstraction-driven MPC controller reduce the value of the cost function $J$, while retaining the certified bound $\lambda$ on the satisfaction probability?
\end{enumerate}
To construct finite-state IMDP abstractions, we extend the IMDP abstraction toolbox from~\cite{Badings2025CDC} to generate $L_\infty$-balls around each action with a radius of $\epsilon$.
We use the model checker Storm~\cite{DBLP:journals/sttt/HenselJKQV22} to compute optimal policies for IMDPs, and Gurobi~\cite{gurobi} to solve the MPC optimization problems. 
For the MPC controller, we first construct a general optimization problem offline based on the PWA approximation of the nonlinear system and the logic-driven selection of the $L_p$-balls, which we store in memory. 
Then, at every step $k \in \mathbb{N}$, we instantiate this general problem with the current state $x_k$ and define the cost function, which provides the MPC problem to solve at the current time-step.
All experiments are run on a laptop with an Intel i7-1185G7 CPU and 16~GB of RAM. 
The code to reproduce the experiments is available in a long-term access repository~\cite{riccardi_CodePublicationTemporal_2026} and on GitHub.\footnote{\tiny \tt{https://github.com/alessandro-riccardi/abstraction-mpc-integration}}

\subsection{Benchmarks and IMDP abstractions}
We consider (I)~a double integrator, (II)~mountain car~\cite{towers_GymnasiumStandardInterface_2025}, and (III) the Dubins car from in Example~\ref{ex:dubins}.

The dynamics for the double integrator are defined as
\begin{align}
    & x_{k+1} = \begin{bmatrix}
        1 & \tau \\ 0 & 1
    \end{bmatrix} x_{k} + \begin{bmatrix}
        \frac{\tau^2}{2} \\ \tau
    \end{bmatrix} u_k + w_k,
\end{align}
with sampling time $\tau=1$, stochastic noise $w_k \sim \mathcal{N}(0, 0.15 I_2)$, state space ${\mathcal{X}} = {\mathbb{R}}^2$, and constrained input $u_k \in {\mathcal{U}} = [-5,5]$. 
The reach-avoid specification is to reach ${\mathcal{X}}_G = [-4,4] \times [-2,2]$ without leaving ${\mathcal{X}}_U = [-21,21] \times [-21,21]$.
We partition the state space into $21 \times 21 = 441$ regions and define $|{\mathcal{A}}| = 21$ abstract actions, each associated with an $L_\infty$-ball with the centers $u_1,\ldots,u_{21}$ forming a uniform gridding of ${\mathcal{U}}$ and with a fixed radius $\epsilon$ (as specified in Sec.~\ref{subsec:criteria_balls}).

The dynamics for the mountain car are taken from~\cite{towers_GymnasiumStandardInterface_2025}:
\begin{align}
    & v_{k+1} = v_k + \tau \left(P u_k - g \cos( p_k)\right) + w_k^{[2]},\\
    & p_{k+1} = p_{k} + \tau ( v_{k+1} - w_k^{[2]}) + w_k^{[1]},
\end{align}
where $p_k\in{\mathbb{R}}$ is the position and $v_k\in[-0.07,0.07]$ the velocity, $\tau = 2$, $P = 0.0015$, and $g = 0.0025$.
The noise is distributed as $w_k \sim \mathcal{N}(0, \text{diag}(0.005,0.0005))$, and the input is contrained as $u_k \in {\mathcal{U}} = [-1,1]$.
The reach-avoid specification is to reach ${\mathcal{X}}_G = [0.45,0.6] \times [-0.07, 0.07]$ without reaching a position $p_k$ outside $[-1.2,0.6]$.
We partition ${\mathcal{X}}$ into $360\times 140$ cells and define $|{\mathcal{A}}|=5$ actions, each associated with an $L_\infty$-ball with uniform gridding of ${\mathcal{U}}$ and fixed radius $\epsilon$ (as specified in  Sec.~\ref{subsec:criteria_balls}).

\subsection{IMDP abstraction and selection of $\epsilon$} \label{subsec:criteria_balls}

To answer Q1, we generate IMDP abstractions (as described in Sec.~\ref{sec:Interface} and \ref{sec:IMDP_abstraction}) for the values of $\epsilon$ in the first column of Table~\ref{tab:simulations-results}.
For each IMDP, we compute an optimal policy ${\sigma}^\star$ as defined in \eqref{eq:optimal_policy} and choose the threshold~$\lambda$ as the certified lower bound on the satisfaction probability, i.e., $\lambda = \min_{P \in {\mathcal{P}}} {\mathbb{P}}_{\mathbf{M}}^{{\sigma},P}({\mathbf{Y}})$ for the initial states $x_0$ shown in Figs.~\ref{fig:integrator-05}-\ref{fig:dubins_simulation}.

\begin{figure}[t!]%
    \centering
    \includegraphics[width=.95\linewidth]{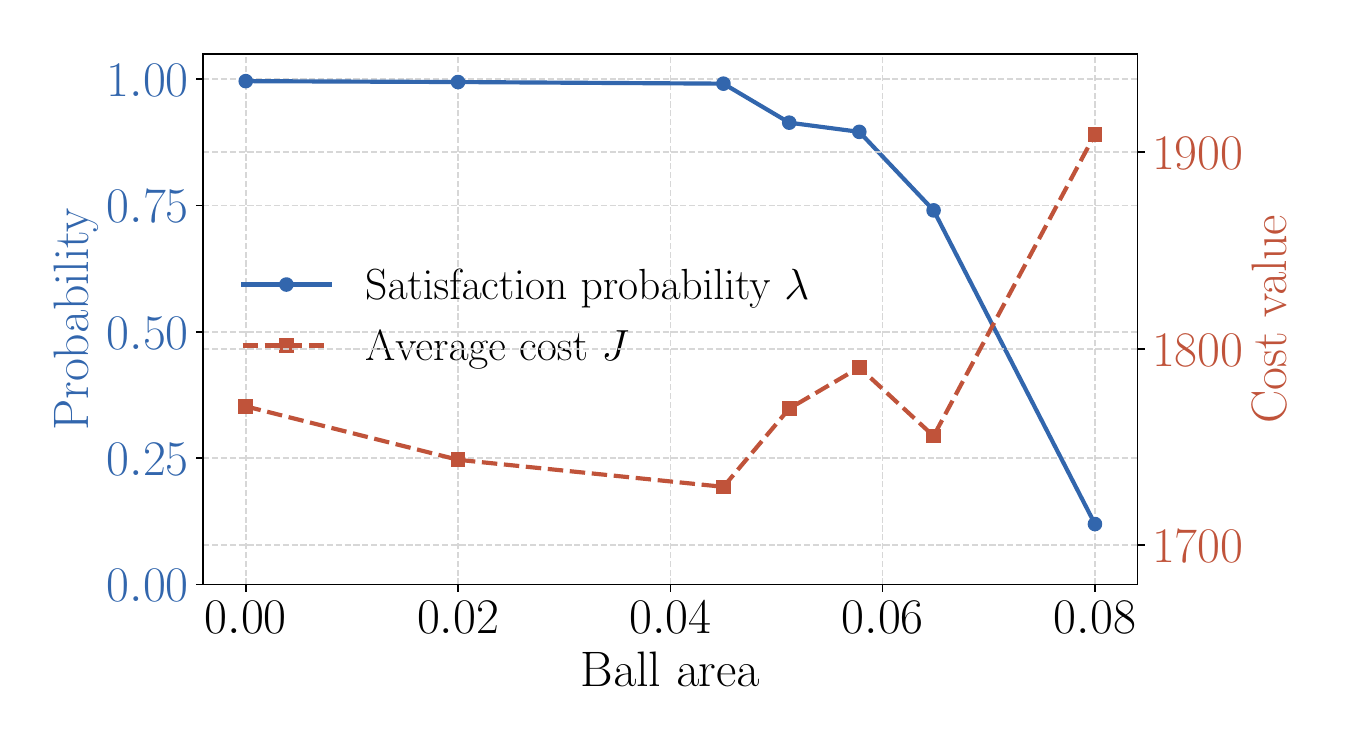}
    \caption{Change in the satisfaction probability of the control specification and of the online MPC performance as a function of the area of the $L_\infty$-balls. Further increasing the size above the elbow value will sharply decrease the satisfaction probability. The value of the cost function $J$ will instead decrease until the elbow point in $\lambda$ (i.e., online MPC performance will increase) and then it will start to increase rapidly. Thus, in this case, the best selection of $\epsilon$ occurs exactly at the elbow: greatest performance improvement with the smallest guarantee reduction. }
    \label{fig:dual_axis}%
\end{figure}

The results in Table~\ref{tab:simulations-results} show that by increasing the value of $\epsilon$---i.e., by enlarging the $L_\infty$-ball ${\mathcal{F}_\textnormal{set}}(x, \sigma(s_i)) = {B}(u_i, \epsilon)$---the value of $\lambda$ decreases, but not in a linear fashion.
In particular, we observe that increasing $\epsilon$ will slowly decrease $\lambda$ until an \emph{``elbow''} point where a sharp decrease occurs, as shown in Fig.~\ref{fig:dual_axis} for the Dubins car: if the area of each $L_{\infty}$-ball increase above 0.045, then $\lambda$ starts decreasing sharply.\footnote{We report the $L_{\infty}$-ball's area instead of the radius, as the radius differs between the state space dimensions; see Tab.~\ref{tab:simulations-results}.} The same phenomenon is observed for the other benchmarks reported in Tab.~\ref{tab:simulations-results}. We select this elbow point as the best trade-off for the selection of $\epsilon$, giving the maximum optimization area for the MPC for a small loss in $\lambda$: in the case of the Dubins car the reduction is of the $0.5\%$ at $L_{\infty}$-ball area $0.045$. 

\begin{figure}[t!]
    \centering
    \begin{subfigure}[t]{0.45\linewidth}
        \includegraphics[width=\linewidth]{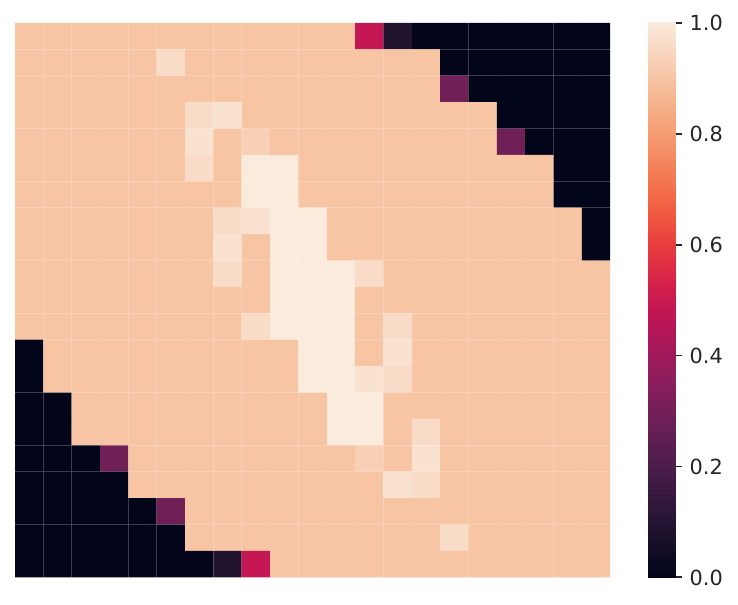}
        \caption{$\epsilon = 0.3$}
    \end{subfigure} \hfill
    \begin{subfigure}[t]{0.45\linewidth}
        \includegraphics[width=\linewidth]{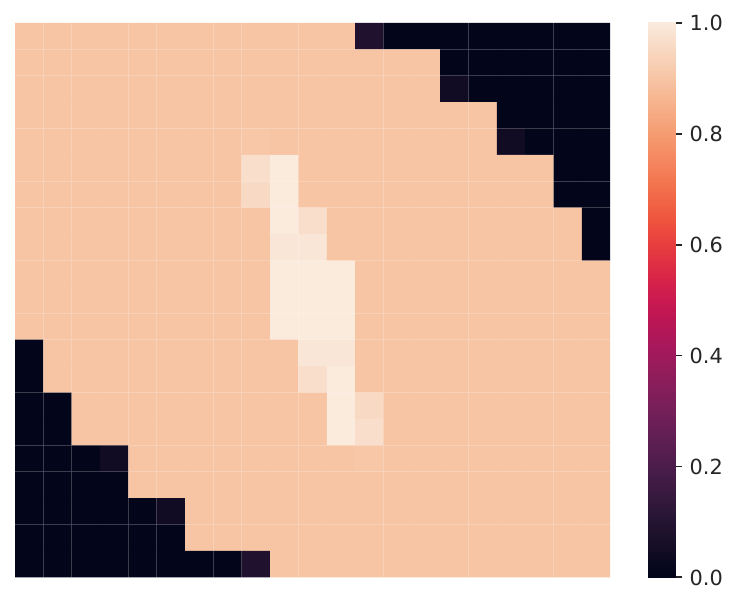}
        \caption{$\epsilon = 0.5$}
    \end{subfigure} 
    \caption{Heat maps of the lower bound of the satisfaction probability for the double integrator.}
    \label{fig:heat-maps-integrators}
\medskip
    \centering
    \begin{subfigure}[t]{0.45\linewidth}
        \includegraphics[width=\linewidth]{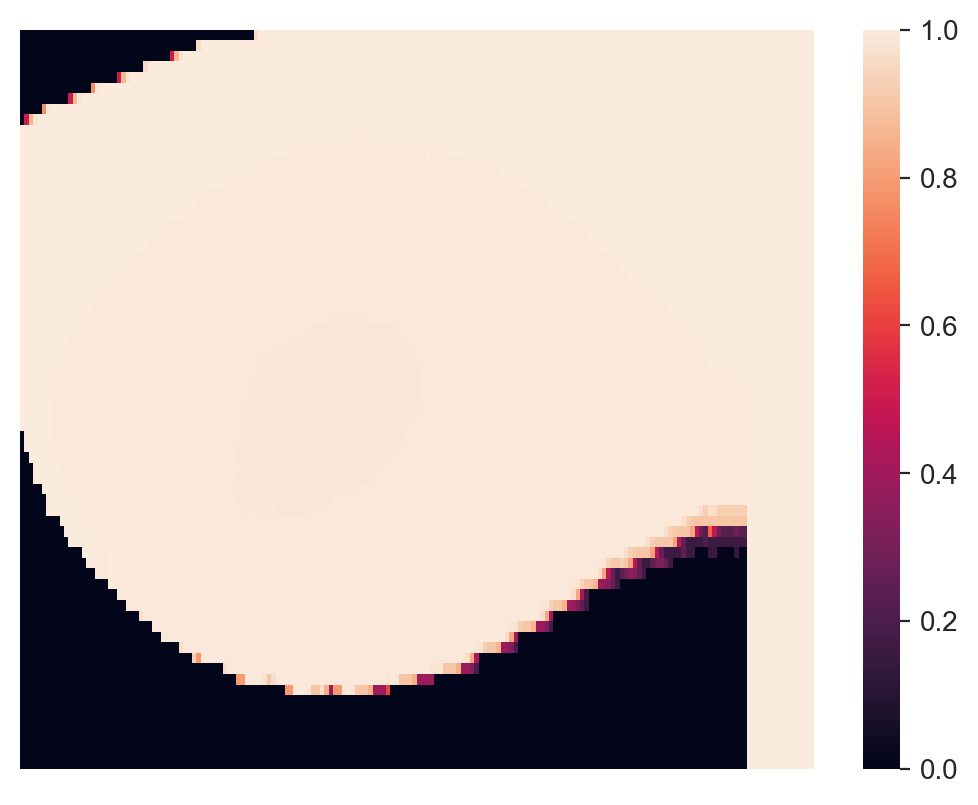}
        \caption{$\epsilon = 0.1$}
    \end{subfigure} \hfill
    \begin{subfigure}[t]{0.45\linewidth}
        \includegraphics[width=\linewidth]{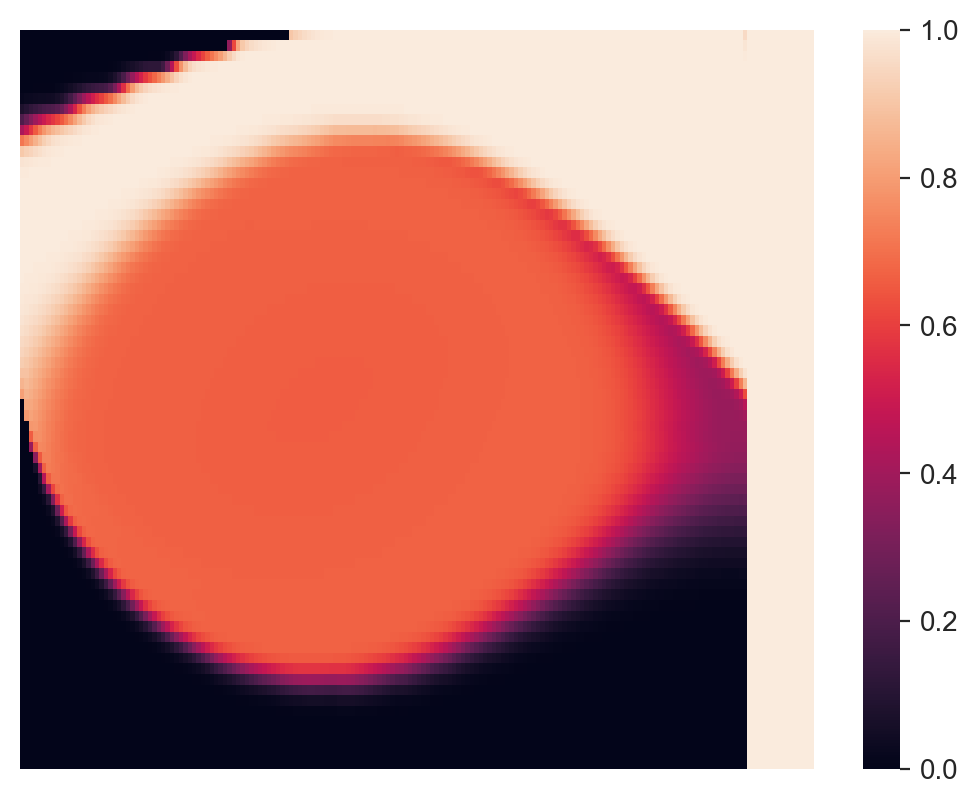}
        \caption{$\epsilon = 0.2$}
    \end{subfigure} 
    \caption{Heat maps of the lower bound of the satisfaction probability for the mountain car.}
    \label{fig:heat-maps-mountain-car}
\medskip
    \centering
    \begin{subfigure}[t]{0.45\linewidth}
        \includegraphics[width=\linewidth]{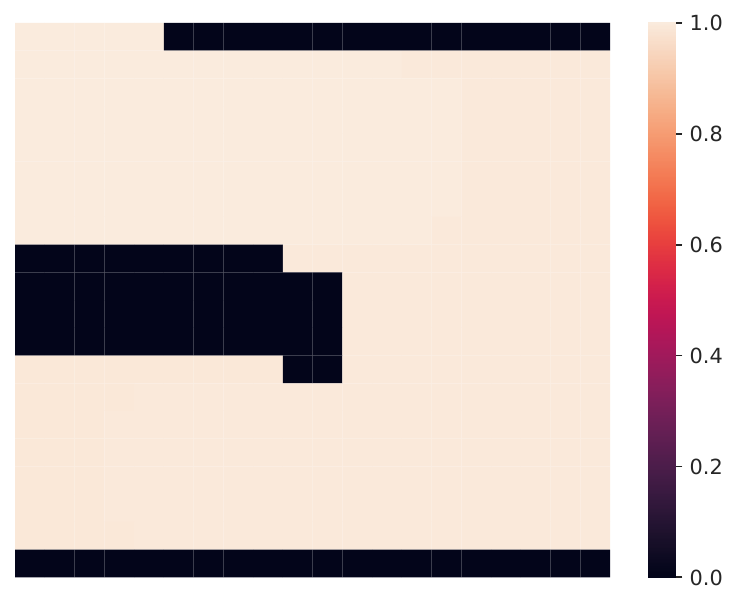}
        \caption{$\epsilon = [0.15;0.3]$}
    \end{subfigure} \hfill
    \begin{subfigure}[t]{0.45\linewidth}
        \includegraphics[width=\linewidth]{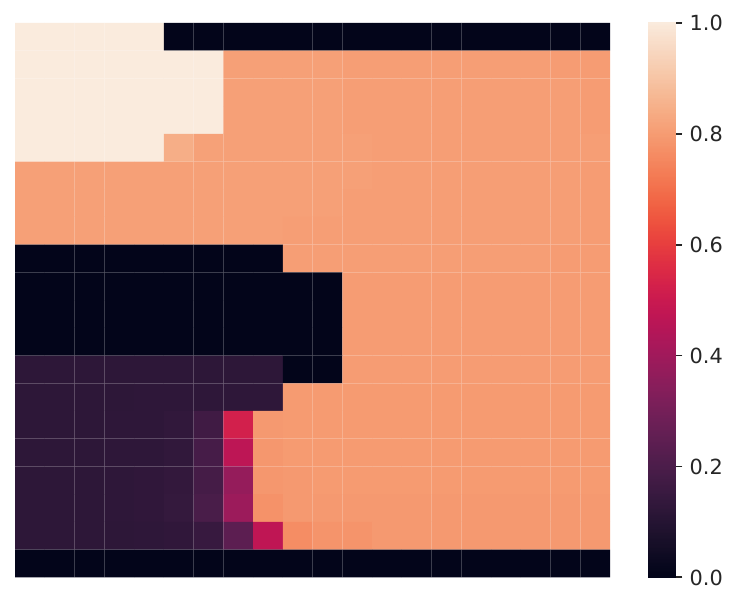}
        \caption{$\epsilon = [0.2;0.4]$}
    \end{subfigure} 
    \caption{Heat maps of the lower bound of the satisfaction probability for the Dubins car.}
    \label{fig:heat-maps-dubins}
\end{figure}

The heat maps in Fig.~\ref{fig:heat-maps-integrators}--\ref{fig:heat-maps-dubins} show the lower bound on the satisfaction probability $\lambda$ for \emph{any} initial state $x_0 \in {\mathcal{X}}$, for the different benchmarks and values of $\epsilon$.
These figures confirm that a higher $\epsilon$ generally decreases the lower bound on the satisfaction probability, but by how much depends strongly on the particular benchmark.

\subsection{Simulation results}
\label{subsec:simulation_results}
To answer Q2, we run simulations under the MPC controller.
For each benchmark and value of $\epsilon$, we run 100 simulations of the closed-loop (stochastic) system under the MPC controller.
The performance for $\epsilon=0$ is equivalent to a vanilla IMDP abstraction (equivalent to~\cite{DBLP:journals/tac/LahijanianAB15,Badings2025CDC}) and is thus the baseline to which we compare our performance.
For a fair comparison, we use the same sequence of noise realizations $w_k$ for the different values of $\epsilon$.

\begin{figure}[t!]
    \centering
    \begin{subfigure}[t]{0.48\linewidth}
        \includegraphics[width=\linewidth]{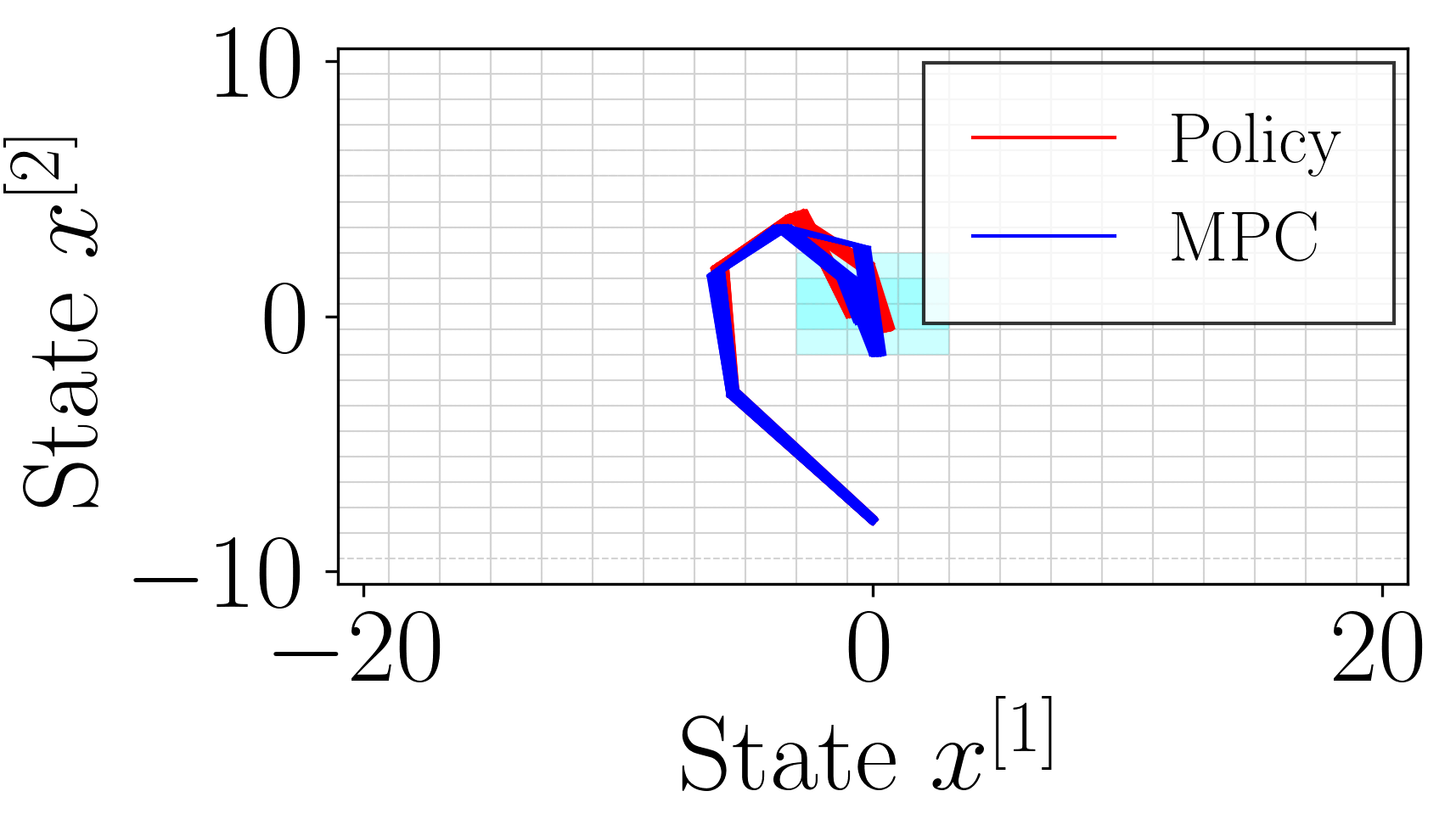}
        \caption{$N=3$, $\epsilon = 0,3$}
    \end{subfigure} 
    \begin{subfigure}[t]{0.48\linewidth}
        \includegraphics[width=\linewidth]{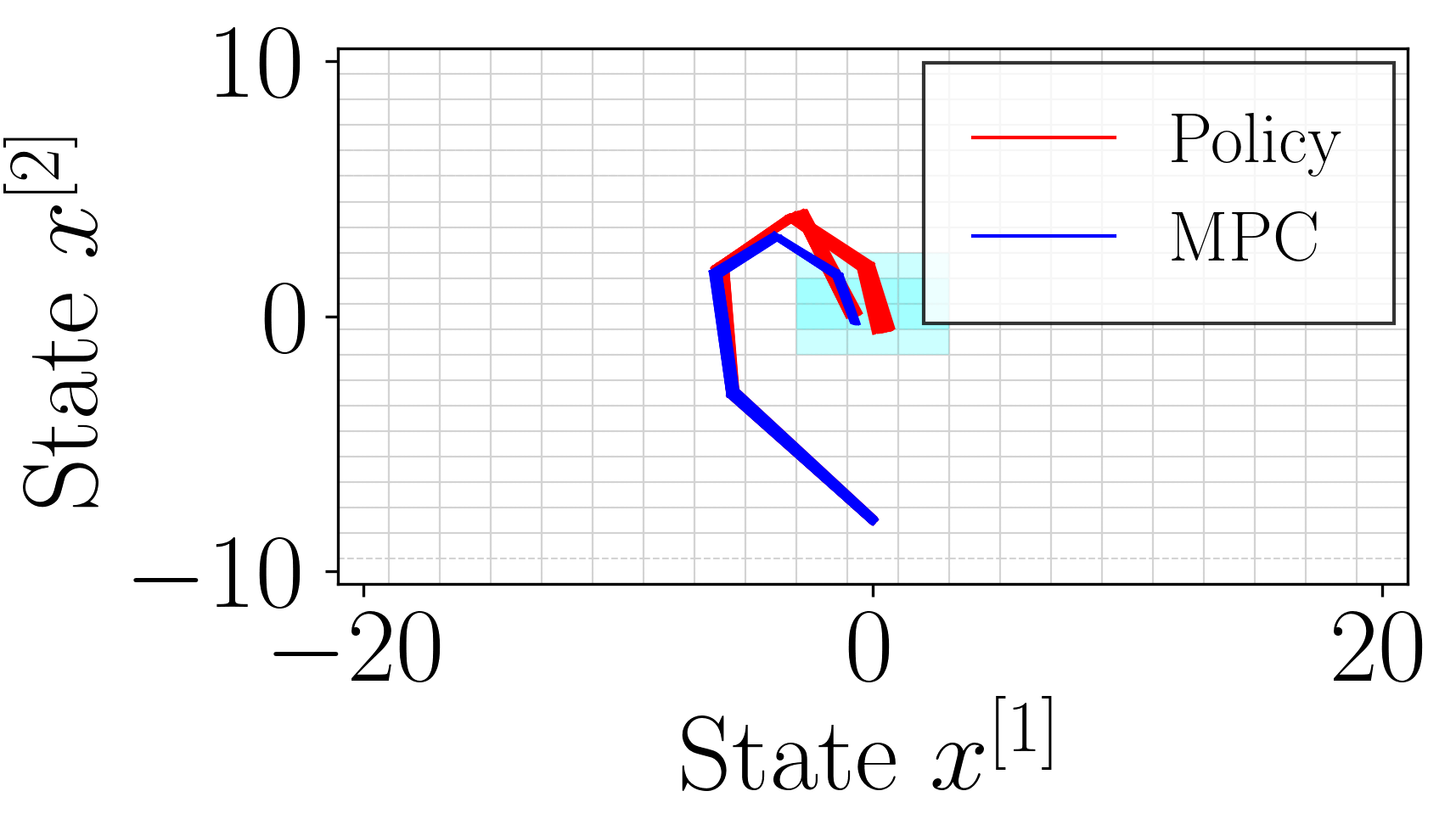}
        \caption{$N=5$, $\epsilon = 0,5$}
    \end{subfigure}
    \caption{Simulated state trajectories for the double integrator, showing the baseline without MPC (policy) and the MPC.}
    \label{fig:integrator-05}
\medskip
    \centering
    \begin{subfigure}[t]{0.48\linewidth}
        \includegraphics[width=\linewidth]{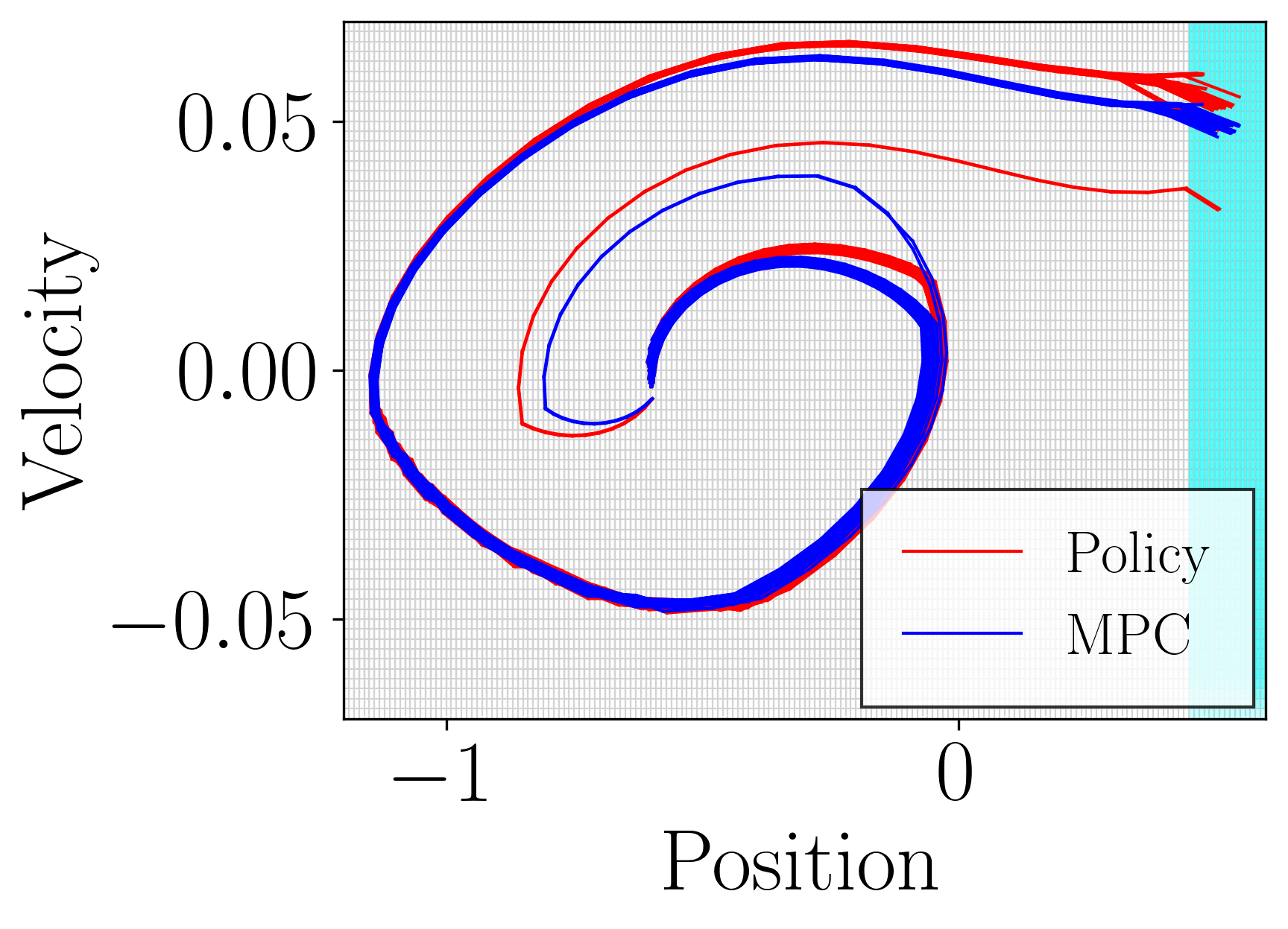}
        \caption{$\epsilon=0.1$}
    \end{subfigure} 
    \begin{subfigure}[t]{0.48\linewidth}
        \includegraphics[width=\linewidth]{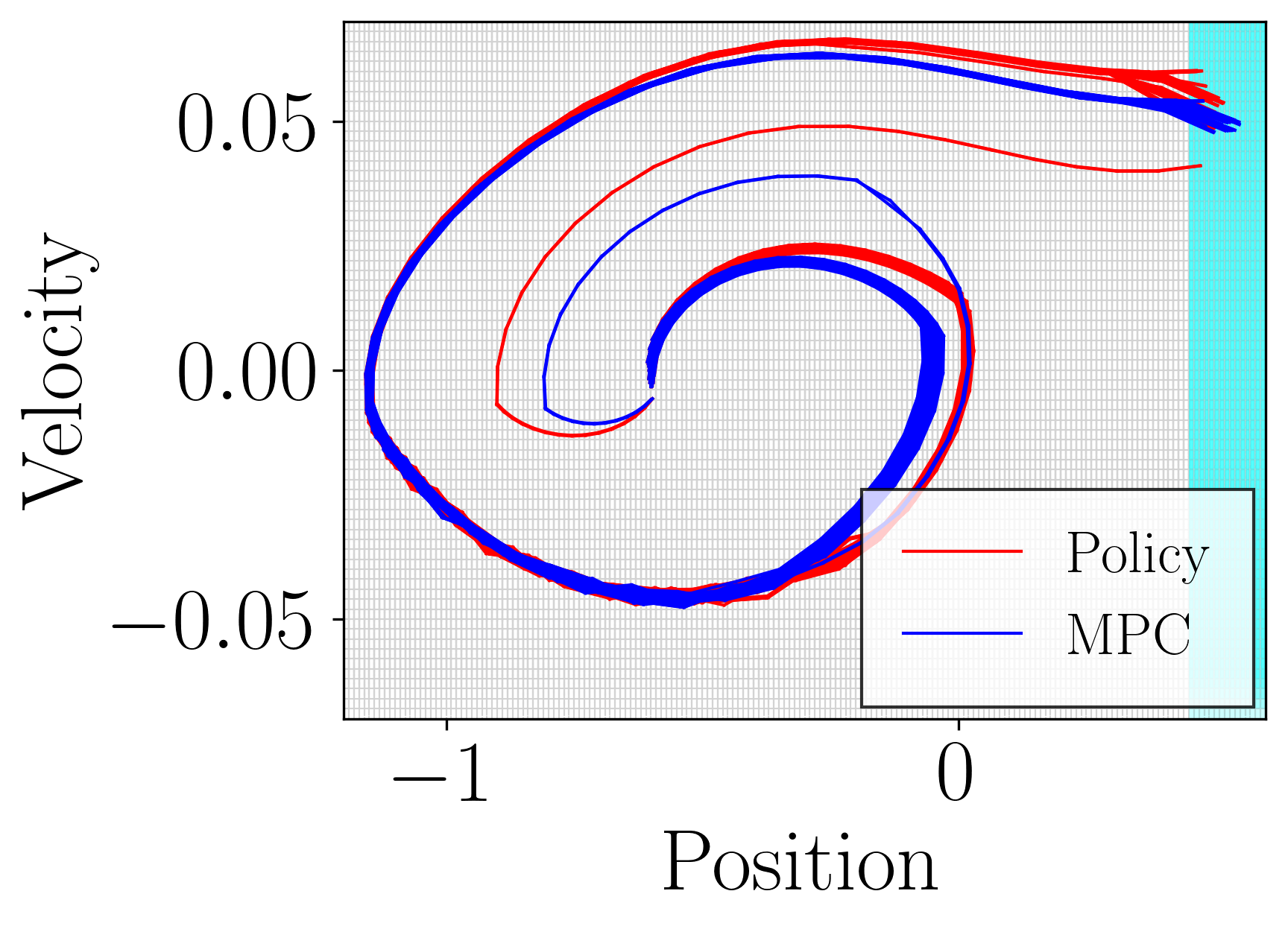}
        \caption{$\epsilon=0.2$}
    \end{subfigure}
    \caption{Mountain car simulations with different $\epsilon$ for the baseline (policy; red) and our MPC controller (blue).
    }
    \label{fig:mountain-car}
\medskip
    \centering
    \begin{subfigure}[t]{0.48\linewidth}
        \includegraphics[width=\linewidth]{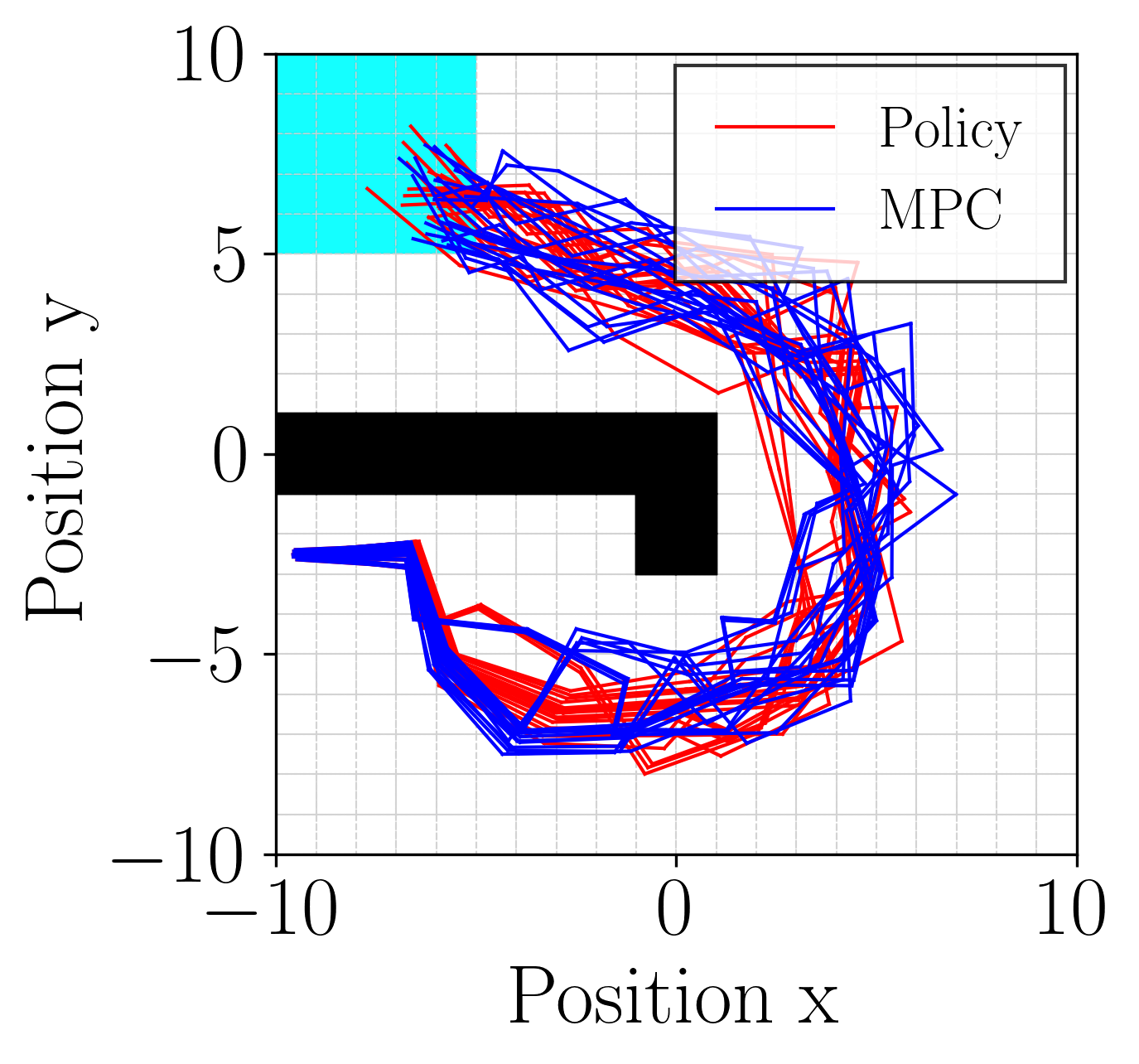}
        \caption{$\epsilon=[0.1;0.2]$}
    \end{subfigure} 
    \begin{subfigure}[t]{0.48\linewidth}
        \includegraphics[width=\linewidth]{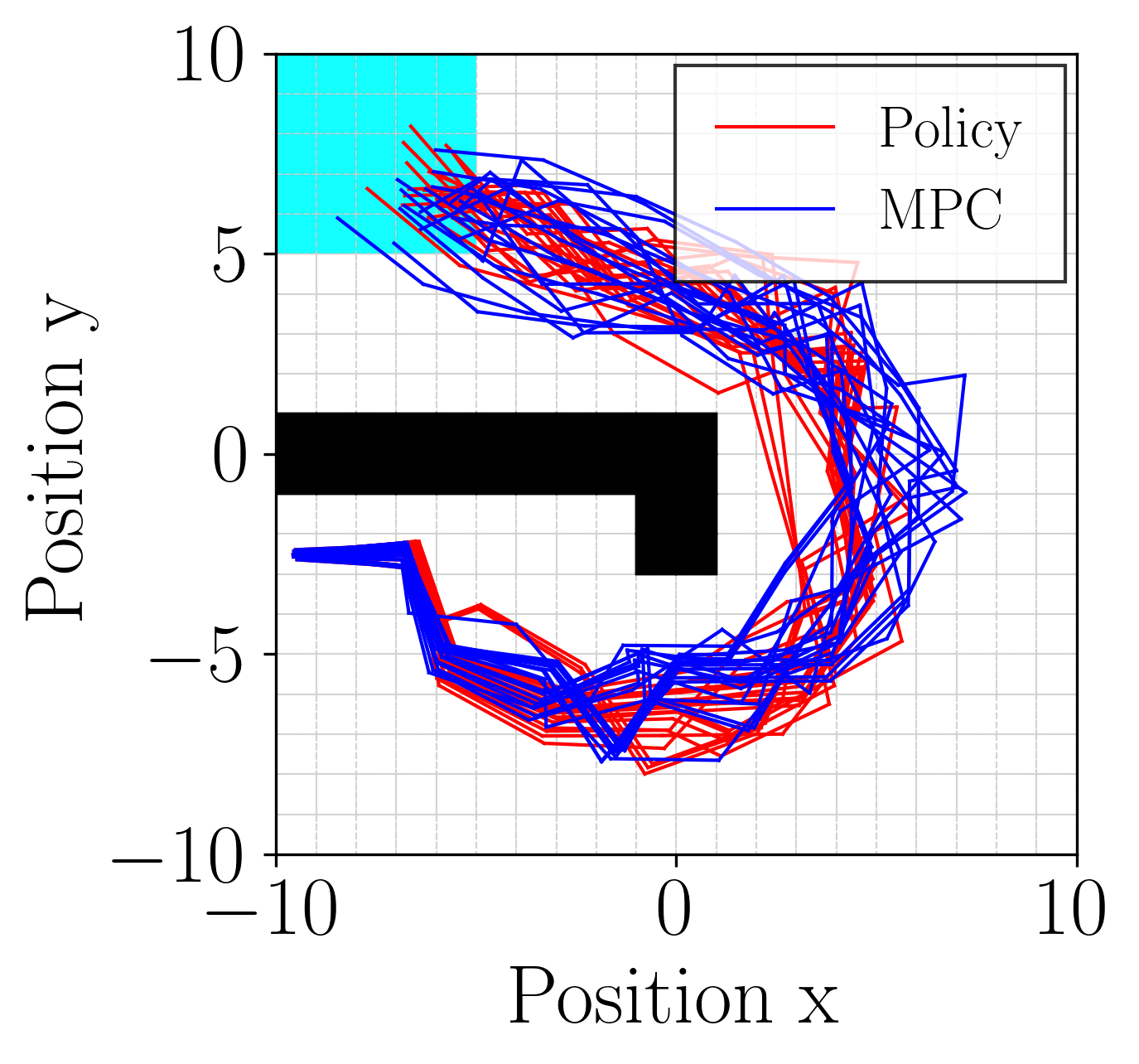}
        \caption{$\epsilon=[0.15;0.3]$}
    \end{subfigure}
    \caption{Dubins simulations with different $\epsilon$ for the baseline without MPC (policy; red) and our MPC controller (blue).
    }
    \label{fig:dubins_simulation}
\end{figure}

For each benchmark and $\epsilon$, Table~\ref{tab:simulations-results} shows the average total cost ${\mathbb{E}}[J]$ (over the 100 simulations), as well as the decomposition into the reference error ${\mathbb{E}}[J_{\textrm{state}}]$ and the control effort ${\mathbb{E}}[J_{\textrm{input}}]$ (such that ${\mathbb{E}}[J] = {\mathbb{E}}[J_{\textrm{state}}] + {\mathbb{E}}[J_{\textrm{input}}]$).
The final two columns show the time required for the abstraction ($\textrm{T}^{\textrm{abs}})$ and the average time required to solve the optimization problem for the MPC at each time step $k$ ($\textrm{T}^{\textrm{MPC}}_{\textrm{step}}$).
Figs.~\ref{fig:integrator-05}--\ref{fig:dubins_simulation}  show simulated state trajectories under both the vanilla abstraction policy, versus the MPC controller.

\textbf{Double integrator\,\,} 
We consider two different prediction horizons $N=3, 5$ and use weighting matrices $Q=\textrm{diag}(1,1)$, $R=1$. 
The offline time required to construct the MPC model is approximately 28 seconds for all cases. 
For both $N=3,5$, the best performance is achieved for $\epsilon = 0.5$, which reduces the cost by $11.6\%$ at a loss in $\lambda$ of around $10\%$. 
Furthermore, the cost reduction is slightly higher for $N=5$ than $N=3$, showing that a longer horizon indeed allows for better online optimization, at the cost of higher time ($\textrm{T}^{\textrm{MPC}}_{\textrm{step}}$) to solve each MPC instance.

\AddToHook{env/tabular/begin}{\scriptsize}
\setlength{\tabcolsep}{3pt}
\begin{table}\centering 
\caption{Experimental results for all benchmarks, showing the radius for the $L_p$-balls ($\epsilon$), the certified lower bound on the satisfaction probability ($\lambda$), the average cost per trajectory, and the time in seconds to generate the abstraction ($T^\text{abs}$) and solve the MPC problem per step ($T^{\text{MPC}}_{\text{step}}$).}
\label{tab:simulations-results}
\renewcommand{\arraystretch}{1.1}
\begin{tabular}{@{}l|l|lll|ll@{}}
\toprule
\multicolumn{7}{c}{\textbf{Double Integrator} with $N=3$}\\
\midrule
$\epsilon$ & $\lambda$ & ${\mathbb{E}}[J]$ & ${\mathbb{E}}[J_{\textrm{state}}]$ & ${\mathbb{E}}[J_{\textrm{input}}]$ & $\textrm{T}^{\textrm{abs}}$ & $\textrm{T}^{\textrm{MPC}}_{\textrm{step}}$ \\
0  & 0.999 & 141.21 & 74.32 & 66.90 &  6.53 & --\\
0.1& 0.992 &   137.42  & 73.48 & 63.93 &   6.99 & 0.23 \\
0.3& 0.892 &   137.61 & 74.99 & 62.62 &   9.23 & 0.18 \\
0.5& 0.898 &  126.58 & 71.05 & 55.53 &   11.92 & 0.18\\
\midrule
\multicolumn{7}{c}{\textbf{Double Integrator} with $N=5$}\\
\midrule
$\epsilon$ & $\lambda$ & ${\mathbb{E}}[J]$ & ${\mathbb{E}}[J_{\textrm{state}}]$ & ${\mathbb{E}}[J_{\textrm{input}}]$ & $\textrm{T}^{\textrm{abs}}$ & $\textrm{T}^{\textrm{MPC}}_{\textrm{step}}$ \\
0 & 0.999 &  141.21 & 74.32 & 66.90 & 6.53 & --\\
0.1& 0.992 &  137.49 & 73.17 & 64.31 &   6.99 & 0.53 \\
0.3&  0.892 & 136.84  & 74.48 & 62.37 &   9.23 & 0.74 \\
0.5& 0.898 & 124.79 & 70.38 & 54.41 &  11.92 & 0.74\\
\midrule
\multicolumn{7}{c}{\textbf{Mountain Car}}\\
\midrule
$\epsilon$ & $\lambda$ & ${\mathbb{E}}[J]$ & ${\mathbb{E}}[J_{\textrm{state}}]$ & ${\mathbb{E}}[J_{\textrm{input}}]$ & $\textrm{T}^{\textrm{abs}}$ & $\textrm{T}^{\textrm{MPC}}_{\textrm{step}}$ \\
0  & 0.991 & 192.80 & 107.42 & 85.38 &   743.69 & -- \\
0.1 & 0.987 &  90.96 & 58.02 & 32.94 &   750.12 & 7.05 \\
0.15& 0.964 &  91.82 & 59.02 & 32.80  & 773.65 & 7.04 \\
0.2 & 0.744 &  92.38 & 59.14 & 33.23 &  904.07 & 7.07\\
\midrule
\multicolumn{7}{c}{\textbf{Dubins Car}}\\
\midrule
$\epsilon$ & $\lambda$ & ${\mathbb{E}}[J]$ & ${\mathbb{E}}[J_{\textrm{state}}]$ & ${\mathbb{E}}[J_{\textrm{input}}]$ & $\textrm{T}^{\textrm{abs}}$ & $\textrm{T}^{\textrm{MPC}}_{\textrm{step}}$ \\
$[0;0]$  & 0.997 &  1770.68 & 1642.67 & 128.01 &  18.76 & -- \\ 
$[0.1;0.2]$& 0.995&  1743.48 & 1622.85 & 120.63 &   21.51 & 4.48 \\
$[0.15;0.3]$& 0.992 &  1729.71 & 1614.11 & 115.60 &   31.37 & 5.08 \\
$[0.2;0.4]$& 0.121 &  1909.15 & 1790.50 & 118.65 &  2522.53 & 5.29 \\
\bottomrule
\end{tabular}
\end{table}

\textbf{Mountain car\,\,}
We consider an MPC horizon of $N=3$ and weighting matrices $Q=\textrm{diag}(1,0)$, $R=1$.
The offline time to construct the general MPC problem is around 6 minutes. In this case, the optimization shows significant performance improvement ($52.8\%$) for $\epsilon = 0.1$, with a minor loss in $\lambda$ of $0.45\%$. Of particular relevance is the improvement in the control effort, with a gain of $61.4\%$, which can have a huge impact on the energy requirements to execute the control action and satisfy the specification. 

\textbf{Dubins car\,\,}
We consider an MPC horizon $N=3$ and weighting matrices $Q=\textrm{diag}(1,1,0)$, $R=\textrm{diag}(1,1)$. The offline time required to construct the general MPC problem is approximately 3 minutes.
Also in this case, the integrated abstraction-MPC controller shows performance improvements both in the state and in the control action. 
The best results are obtained for the configuration $\epsilon = [0.15; 0.3]$ (two values are required because the system has two inputs), where a $\lambda$-loss of the $0.47\%$, we obtain a $1.73\%$ improvement in the cost for the state, and a $9.7\%$ improvement in the control effort. 

\begin{remark}
    In practice, the complexity of the MIQP for the MPC problem grows exponentially with the number of cells in the state space partition.
    Thus, constructing a single MIQP for the entire problem can be prohibitive if the abstraction is too fine-grained.
    Alternatively, we can consider a different `general' problem for each cell of the partition, thus constructing a set of $N_s$ smaller problems, which are then updated with constraints accounting for the measured state. Constructing these local problems requires computing the subset of reachable cells within the prediction horizon. Such an approach can significantly improve online computational performance at the cost of a much longer offline abstraction time and increased memory usage to store one optimization subproblem for each abstraction cell. 
\end{remark}

\section{Conclusions}
\label{sec:Conclusion}
We have presented a novel policy synthesis framework for discrete-time stochastic systems that integrates {offline} abstraction with online model predictive control (MPC).
In the abstract model, which we represented as an interval Markov decision process (IMDP), each action is associated with a set of control inputs for the dynamical system. 
By performing robust value iteration on the abstract IMDP, we obtain a set of policies for the dynamical system, each of which satisfies a given logical specification with a certified minimum probability.
Online, we use MPC to further optimize a desired cost function (e.g., total control effort) by choosing inputs within this certified set of policies from the abstraction.
The resulting MPC controller optimizes the cost function while retaining the certified lower bound on the satisfaction probability from the abstraction.
Our experiments demonstrated that our approach yields better control performance than vanilla abstraction techniques, with a tunable and often only small decrease in the guaranteed probability of satisfying the specification. 

To further improve performance, we wish to explore adaptive abstraction schemes that use a variable $\epsilon$ (i.e., the radius of the $L_p$-balls to define actions) across the state space.
To improve the tightness of the abstract model, we will investigate using models other than IMDPs to represent the abstraction, e.g., as done in~\cite{DBLP:conf/hybrid/MathiesenHL25,DBLP:journals/corr/abs-2507-02213}.
Finally, we will explore methodologies to improve the online computation times associated with the solution of the mixed-integer MPC problems introduced in the control scheme.

\begin{ack}                             
This research has been supported by the EPSRC grant EP/Y028872/1, Mathematical Foundations of Intelligence: An ''Erlangen Programme`` for AI, and by the European Research Council (ERC) under the European Union's Horizon 2020 research and innovation program (Grant agreement No. 101018826) -- Project CLariNet. 
\end{ack}

\appendix

\section{Proof of Thm.~\ref{thm:PASR_synthesis}}
\label{appendix:PASR_synthesis}
As in~\cite{DBLP:journals/tac/HaesaertS21}, the satisfaction probability ${\mathbb{P}}^{\hat{\pi}}_{{\mathbf{S}}}({\mathbf{Y}})$ under a fixed policy $\hat{\pi} \colon {\mathcal{X}} \to {\mathcal{U}}$ can be computed via a dynamic programming recursion on the value function $V^{\hat{\pi}}_{k} \colon {\mathcal{X}} \to {\mathbb{R}}$, $k\in{\mathbb{N}}$, where
\begin{align}
    \label{eq:thm:recursionV}
    V_{k+1}^{\hat{\pi}}(x) = \int_{{\mathcal{X}}} \max\left\{ \mathbbold{1}_{{\mathcal{X}}_F}(\xi), V_{k}^{\hat{\pi}}(\xi) \right\} \cdot {\mathbf{t}}(\mathrm{d}\xi \mid x,\hat{\pi}(x)),
\end{align} 
where ${\mathcal{X}}_F \subseteq {\mathcal{X}}$ is the set of states from which ${\mathcal{Y}}$ is satisfied with probability one.
Let $V_\star^{\hat{\pi}}$ be the least fixed point of this recursion.
The satisfaction probability is given as ${\mathbb{P}}^{\hat{\pi}}_{{\mathbf{S}}}({\mathbf{Y}}) = \int_{\mathcal{X}} V_\star^{\hat{\pi}}(\xi) {\mu_{x_0}}(\mathrm{d}\xi)$.

Due to the monotonicity of the value function, the value function for any policy ${\pi}$ that satisfies \eqref{eq:PASR_policy} is bounded as $\check{V}_\star(x) \leq V_\star^{\hat{\pi}}(x) \leq \hat{V}_\star(x)$ for all $x \in {\mathcal{X}}$, where $\check{V}_\star(x)$ and $\hat{V}_\star(x)$ are the fixed points of the recursions
\begin{equation*}
    \begin{split}
        \check{V}_{k+1}(x) = \min_{u \in {\mathcal{F}_\textnormal{set}}(x,\sigma)}
    \int_{{\mathcal{X}}} \max\left\{ \mathbbold{1}_{{\mathcal{X}}_F}(\xi), \check{V}_{k}(\xi) \right\} \\ \cdot \, {\mathbf{t}}(\mathrm{d}\xi \mid x,\hat{\pi}(x)),
    \end{split}
\end{equation*}
\begin{equation*}
    \begin{split}
        \hat{V}_{k+1}(x) = \max_{u \in {\mathcal{F}_\textnormal{set}}(x,\sigma)}
    \int_{{\mathcal{X}}} \max\left\{ \mathbbold{1}_{{\mathcal{X}}_F}(\xi), \hat{V}_{k}(\xi) \right\} \\ \cdot \, {\mathbf{t}}(\mathrm{d}\xi \mid x,\hat{\pi}(x)),
    \end{split}
\end{equation*}
with the shorthand notation 
\begin{equation*}
    {\mathcal{F}_\textnormal{set}}(x,\sigma) \coloneqq {\mathcal{F}_\textnormal{set}}(x, {\sigma}({\mathcal{R}}(x))).
\end{equation*}
To prove \eqref{eq:PASR_synthesis}, we will show that $\check{V}_{k+1}$ and $\hat{V}_{k+1}$ can be lower bounded by dynamic programming recursions on the IMDP.
For the lower bound, define the following recursion for the IMDP:
\begin{align}
    \label{eq:thm:recursionW}
    \check{W}_{k+1}^{\sigma}(s) = \min_{\nu \in {\mathcal{P}}(s,\sigma(s))} \sum_{s' \in {\mathcal{S}}} \max\{ \mathbbold{1}_{{\mathcal{S}}_F}, W_{k}^{{\sigma}}(s') \} \cdot \nu(s').
\end{align}
The minimum satisfaction probability $\min_{P \in {\mathcal{P}}} {\mathbb{P}}_{\mathbf{M}}^{{\sigma},P}({\mathbf{Y}})$ for the IMDP is computed as $\sum_{s \in {\mathcal{S}}} {\mu_{s_0}}(s) \cdot \check{W}_\star^{{\sigma}}(s)$, where $\check{W}_\star^{{\sigma}}$ is the least fixed point of the recursion in \eqref{eq:thm:recursionW}.
Let us show that, if $\check{W}_{k}^{\sigma}(s) \leq \check{V}_{k}(x)$ for all $(x,s) \in {\mathcal{R}}$, then also $\check{W}_{k+1}^{\sigma}(s') \leq \check{V}_{k+1}(x')$ for all $(x,s) \in {\mathcal{R}}$.
First, since ${\mathcal{R}}$ models a partition of ${\mathcal{X}}$, the integral over ${\mathcal{X}}$ can be decomposed as the sum over the IMDP states $s' \in {\mathcal{S}}$:
\begin{equation*}
    \begin{split}
        \check{V}_{k+1}(x) =
        \!\!\!\min_{u \in {\mathcal{F}_\textnormal{set}}(x,\sigma)} \sum_{s' \in S}
        \int_{{{\mathcal{R}}^{{-1}}}(s')}\!\!\!\max\left\{ \mathbbold{1}_{{\mathcal{X}}_F}(\xi), \check{V}_{k}(\xi) \right\} \\ \cdot \, {\mathbf{t}}(\mathrm{d}\xi \mid x,\hat{\pi}(x)).
    \end{split}
\end{equation*}
Next, observe that for all $(x,s) \in {\mathcal{R}}$, we have $\check{V}_{k}(x) \geq \check{W}^\sigma_{k}(s)$.
Furthermore, due to condition (2) of the PASR, it holds that $\mathbbold{1}_{{\mathcal{X}}_F}(x) = \mathbbold{1}_{{\mathcal{S}}_F}(s)$, and due to (3), there exists a distribution $\nu \in {\mathcal{P}}(s,\sigma(s))$ that \emph{underapproximates} the integral of ${\mathbf{t}}(\mathrm{d}\xi \mid x,\hat{\pi}(x))$.
Hence, we obtain
\begin{align*}
    \check{V}_{k+1}(x) 
    \geq&
    \min_{u \in {\mathcal{F}_\textnormal{set}}(x,\sigma)} \sum_{s' \in S}
    \max\left\{ \mathbbold{1}_{{\mathcal{S}}_F}(s'), \check{W}_{k}^\sigma(s') \right\} 
    \\ 
    &\qquad\qquad\qquad\qquad\quad\cdot \int_{{{\mathcal{R}}^{{-1}}}(s')} \!\!\!\! {\mathbf{t}}(\mathrm{d}\xi \mid x,\hat{\pi}(x)),
    \\
    \geq& \min_{\nu \in {\mathcal{P}}(s,\sigma(s))} \sum_{s' \in S}
    \max\left\{ \mathbbold{1}_{{\mathcal{S}}_F}(s'), \check{W}_{k}^\sigma(s') \right\}  \cdot \nu(s') 
    \\ 
    =& \,\, \check{W}_{k+1}^\sigma(s),
\end{align*}
which proves $\check{V}_{k+1}(x) \geq \check{W}_{k+1}^\sigma(s)$ for all $(x,s) \in {\mathcal{R}}$.
The upper bound $\hat{V}_{k+1}^\sigma(x) \leq \hat{W}_{k+1}(s)$ follows analogously and is thus omitted.    
Altogether, it follows that, for all $(x,s) \in {\mathcal{R}}$, the fixed points of these recursions satisfy
\begin{align}
    \label{eq:thm:fixed_point_bounds}
    \check{W}_{\star}^\sigma(s) \leq \check{V}_{\star}(x),
    \qquad
    \hat{V}_{\star}(x) \leq \hat{W}^\sigma_{\star}(s), 
\end{align}
Since the values in the initial states are linear combinations of the values in \eqref{eq:thm:fixed_point_bounds}, the bounds in \eqref{eq:PASR_synthesis} follow.
    
\section{Proof of Thm.~\ref{thm:IMDP_correctness}}
\label{appendix:IMDP_correctness}
Let us show the three conditions in Def.~\ref{def:PASR} are satisfied:
\begin{enumerate}
    \item The initial distribution is defined as ${\mu_{s_0}}(s) = {\mu_{x_0}}({{\mathcal{R}}^{{-1}}}(s))$ for all $s \in {\mathcal{S}}$, so it holds that $({\mu_{x_0}}, {\mu_{s_0}}) \in {{\mathcal{R}}^\mathcal{P}}$.
    \item The partition is label-preserving (Assumption~\ref{assumption:labels}), so ${h_{\mathbf{S}}}(x) = {h_{\mathbf{M}}}(s)$ trivially holds for all $(x,s) \in {\mathcal{R}}$.
    \item For the third condition, pick any kernel $\mu \coloneqq {\mathbf{t}}(\cdot \mid \hat{x}, \hat{u})$ for $\hat{x} \in {\mathcal{X}}$, $s \in {\mathcal{R}}(\hat{x})$, $a \in {\mathcal{A}}$, and $\hat{u} \in {\mathcal{F}_\textnormal{set}}(\hat{x}, a)$.
    We need to show that there exists a distribution $\nu \in \Delta({\mathcal{S}})$ such that $\nu \in {\mathcal{P}}({\mathcal{R}}(\hat{x}),a)$ and $(\mu,\nu) \in {{\mathcal{R}}^\mathcal{P}}$.
    Define $\nu \in \Delta({\mathcal{S}})$ for all $s' \in {\mathcal{S}}$ as
    $
    \nu(s') = {\mathbf{t}}({{\mathcal{R}}^{{-1}}}(s') \mid \hat{x}, \hat{u}).
    $
    Since $\hat{x} \in {{\mathcal{R}}^{{-1}}}(s)$ and $\hat{u} \in {\mathcal{F}_\textnormal{set}}(\hat{x}, a)$, we know that $\nu(s')$ is within the interval bounds computed in \eqref{eq:IMDP:lb} and \eqref{eq:IMDP:ub}.
    Thus, it follows that $\check{P}(s,a,s') \leq \nu(s') \leq \hat{P}(s,a,s')$ for all $s' \in {\mathcal{S}}$, so it indeed holds that $\nu \in {\mathcal{P}}({\mathcal{R}}(\hat{x}),a)$.
    
    What remains to show is that $(\mu,\nu) \in {{\mathcal{R}}^\mathcal{P}}$, i.e., there exists a probability measure $\mathbb{W}$ such that the conditions in Def.~\ref{def:lifting} hold.
    Define this probability measure $\mathbb{W}$ for all $Z \in \mathcal{B}({\mathcal{X}} \times {\mathcal{S}})$~as 
    \[
        \mathbb{W}(Z) = {\mathbf{t}}( \{ x \in {\mathcal{X}} : \exists s \in {\mathcal{S}}, \, (x,s) \in Z \cap {\mathcal{R}} \} \mid \hat{x}, \hat{u} ).
    \]
    It is easily verified that $\mathbb{W}$ is a lifting for $(\mu,\nu)$, i.e., $(\mu,\nu) \in {{\mathcal{R}}^\mathcal{P}}$, as:
    (1) for all $A \in \mathcal{B}({\mathcal{X}})$, it holds that 
    \begin{align*}
        \mathbb{W}(A, {\mathcal{S}}) = & \,\, {\mathbf{t}}( \{ x \in {\mathcal{X}} : \exists s \in {\mathcal{S}}, \\ 
        & \,\, \quad (x,s) \in (A, {\mathcal{S}}) \cap {\mathcal{R}} \} \mid \hat{x}, \hat{u} ) 
        \\
        = & \,\,{\mathbf{t}}( A \mid \hat{x}, \hat{u} ) = \mu(A);
    \end{align*}
    (2) for all $B \in \mathcal{B}({\mathcal{S}})$, it holds that 
    \begin{align*}
        \mathbb{W}({\mathcal{X}}, B)  = & \,\, {\mathbf{t}}( \{ x \in {\mathcal{X}} : \exists s \in {\mathcal{S}}, \\ & \,\, \quad \, (x,s) \in (A, {\mathcal{S}}) \cap {\mathcal{R}} \} \mid \hat{x}, \hat{u} ) 
        \\
        = & \,\, {\mathbf{t}}(\cup_{s \in B} {{\mathcal{R}}^{{-1}}}(s) \mid \hat{x}, \hat{u}) = \nu(B);
    \end{align*}
    (3) $\mathbb{W}({\mathcal{R}}) = {\mathbf{t}}( \{ x \in {\mathcal{X}} : \exists s \in {\mathcal{S}}, \, (x,s) \in {\mathcal{R}} \cap {\mathcal{R}} \} \mid \hat{x}, \hat{u} ) = {\mathbf{t}}( {\mathcal{X}} \mid \hat{x}, \hat{u}) = 1$.
\end{enumerate}
This concludes the proof.

\bibliographystyle{plain}        % Include this if you use bibtex 
{\bibliography{references}}           % and a bib file to produce the 
                                 % bibliography (preferred). The
                                 % correct style is generated by
                                 % Elsevier at the time of printing.

\appendix

\end{document}

%% file: tikz_packages.tex
\usepackage{tikz} 
\usepackage{pgfplots}
\usepgfplotslibrary{external,fillbetween,colorbrewer}
\pgfplotsset{compat=newest}

\DeclareUnicodeCharacter{2212}{−}
\usepgfplotslibrary{external,}
\usetikzlibrary{patterns,shapes.arrows}

\usepackage{colortbl}
\definecolor{LightCyan}{rgb}{0.88,1,1}

\definecolor{red}{rgb}{0.745,0.192,0.102}
\definecolor{darkgreen}{RGB}{34,161,55}
\definecolor{ruhuisstijlrood}{rgb}{0.745,0.192,0.102}
\definecolor{ruhuisstijlzwart}{rgb}{0,0,0}
\definecolor{ruhuisstijlwit}{rgb}{0.98,0.98,0.98}

\usetikzlibrary{patterns,patterns.meta,arrows,arrows.meta,calc,shapes,shadows,decorations.pathmorphing,decorations.pathreplacing,automata,shapes.multipart,positioning,shapes.geometric,fit,circuits,trees,shapes.gates.logic.US,fit, shadows.blur,shapes.symbols,intersections}

%% file: colors.tex
\definecolor{red}{rgb}{0.745,0.192,0.102}
\definecolor{darkgreen}{RGB}{34,161,55}
\definecolor{darkblue}{RGB}{0,33,71}
\definecolor{ruhuisstijlrood}{rgb}{0.745,0.192,0.102}
\definecolor{ruhuisstijlzwart}{rgb}{0,0,0}
\definecolor{ruhuisstijlwit}{rgb}{0.98,0.98,0.98}

\definecolor{darkred}{rgb}{0.560784314,0.125490196,0.0666666667}

\definecolor{color1}{RGB}{55,126,184} % lightblueish
\definecolor{color2}{RGB}{228,26,28} % red
\definecolor{color3}{RGB}{77,175,74} % green
\definecolor{color4}{RGB}{152,78,163} % purple
\definecolor{color5}{RGB}{255,127,0} % orange
\definecolor{color6}{rgb}{0.5, 1.0, 0.83} % aquamarine
\definecolor{color7}{rgb}{1.0, 0.0, 1.0} % magenta
\definecolor{color8}{rgb}{0.66, 0.66, 0.66} % gray

\colorlet{plot1}{color1}
\colorlet{plot2}{color2}
\colorlet{plot3}{color3}
\colorlet{plot4}{color4}
\colorlet{plot5}{color5}

%% file: overview.tex
\scalebox{0.9}{
\begin{tikzpicture}[
    font=\footnotesize,
    node distance=5cm,
    >=stealth,
    line width=0.3mm,
    auto,
    mybox/.style={rectangle, rounded corners, text width=2.4cm, minimum height=0.5cm, text centered, draw=black, text depth=0.85cm, inner sep=0.05cm}
    ]

    \newcommand\xdist{2cm}
    \newcommand\ydist{0.7cm}

    \node (abstraction) [mybox, fill=color1!20] {\textbf{IMDP} \\ \textbf{abstraction}};
    \node (synthesis) [mybox, fill=color1!20, below=0.9cm of abstraction] {\textbf{Policy} \\ \textbf{synthesis}};

   \node (mpc) [mybox, fill=color3!20, below=0.9cm of synthesis, minimum height=1.4cm, text depth=1.25cm] {\textbf{Online MPC} \\ $\min J(\cdot) \, \text{s.t.}$ \\ $u \in \{ \pi(x) \}_{\pi \in \tilde\Pi}$};
    \node (sys) [mybox, fill=color3!20, below=0.2cm of mpc] {\textbf{System} \\ $x^+ = f(x,u,w)$};

    % Input
    \draw[->] ($(abstraction.east) + (0.3,0.0cm)$) -- node[pos=0.0, right, align=left, xshift=0.0cm] {(1) Dynamics $f(x,u,w)$ \\ (2) Specification + \\ \,\,\,\,\,\,\,\, threshold $\lambda$} (abstraction.east);

    \draw[->] ($(mpc.east) + (0.3,1cm)$) -| node[pos=0.0, right, align=left, xshift=0.0cm] {Piecewise affine \\ approximation \\ of $f(x,u,w)$} ($(mpc.north) + (0.5,0)$);

    % Connecting arrows
    \draw[->] (abstraction.south) -- node[pos=0.5, left, align=center] {Finite IMDP \\ abstraction} (synthesis.north);
    \draw[->] (synthesis.south) -- node[pos=0.5, left, align=center] {Set of verified \\ policies $\tilde\Pi$} (mpc.north);
    
    \draw[->] (mpc.east) -- ($(mpc.east) + (0.8,0cm)$) |- node[pos=0.55, below] {Input $u$} (sys.east);
    \draw[->] (sys.west) -| node[pos=0.45, below] {State $x$} ($(mpc.west) + (-0.8,0cm)$) -- (mpc.west);

    \draw[
        decorate,
        decoration={brace, mirror, amplitude=6pt}
    ]
    ($(abstraction.north west) + (-1.4,0.0)$) 
    -- 
    ($(synthesis.south west) + (-1.4,-0.0)$)
    node[midway, above, xshift=-5pt, rotate=90, align=center] {\textbf{Offline verification}};

    \draw[
        decorate,
        decoration={brace, mirror, amplitude=6pt}
    ]
    ($(mpc.north west) + (-1.4,0.0)$) 
    -- 
    ($(sys.south west) + (-1.4,-0.0)$)
    node[midway, above, xshift=-5pt, rotate=90, align=center] {\textbf{Online control}};
\end{tikzpicture}%
}